\documentclass[10pt,twocolumn]{IEEEtran}

\usepackage{graphicx}
\usepackage{amsmath}
\usepackage{amssymb}
\usepackage[caption=false]{subfig}
\usepackage[noadjust]{cite}
\usepackage{float}
\usepackage{algorithm}
\usepackage{bibentry}
\usepackage{balance}
\usepackage{algorithm}
\usepackage{algorithmicx}
\usepackage{algpseudocode}
\usepackage{xcolor}
\usepackage{subfig}

\graphicspath{{img/}}

\begin{document}

\bstctlcite{IEEEexample:BSTcontrol}

\title{Interference Analysis and Mitigation for Massive Access Aerial IoT Considering 3D Antenna Patterns
\thanks{This work is supported in part by the INL Laboratory Directed Research Development (LDRD) Program under DOE Idaho Operations Office Contract DEAC07-05ID14517, and by NSF CNS through award number 1814727. An  earlier version of this work has been published in~\cite{maeng2019interference}.}\thanks{S. J. Maeng, M. A. Deshmukh, \.{I}. G\"{u}ven\c{c}, and H. Dai are with the Department of Electrical and Computer Engineering, North Carolina State University, Raleigh, NC 27606 USA (e-mail: smaeng@ncsu.edu; madeshmu@ncsu.edu; iguvenc@ncsu.edu; hdai@ncsu.edu).}\thanks{A. Bhuyan is with Idaho National Labs, Idaho Falls, ID 83402 USA (e-mail: arupjyoti.bhuyan@inl.gov).}}

\author{\IEEEauthorblockN{Sung Joon Maeng, Mrugen A. Deshmukh, \.{I}smail G\"{u}ven\c{c}, \textit{Senior Member, IEEE}, Huaiyu Dai, \textit{Fellow, IEEE} and  Arupjyoti Bhuyan}}%

\maketitle

\begin{abstract}
%Internet of things (IoT) technology with both ground and aerial sensors has recently attracted major attention for 5G wireless networks. 
Due to dense deployments of  Internet of things (IoT) networks, interference management becomes a critical challenge. With the  proliferation of aerial IoT devices, such as unmanned aerial vehicles (UAVs), interference characteristics in 3D environments will be different than those in the existing terrestrial IoT networks. In this paper, we consider 3D topology IoT networks with a mixture of aerial and terrestrial links, with low-cost cross-dipole antennas at ground nodes and omni-directional antennas at aerial nodes. Considering a massive-access communication scenario, we first derive the statistics of the channel gain at IoT receivers in closed form while taking into account the radiation patterns of both ground and aerial nodes. These are then used to calculate the ergodic achievable rate as a function of the height of the aerial receiver. We propose an interference mitigation scheme that utilizes 3D antenna radiation pattern with different dipole antenna settings. Our results show that using the proposed scheme, the ergodic achievable rate improves as the height of aerial receivers increases. In addition, the ratio between the ground and aerial receivers that maximizes the peak rate also increases with the aerial IoT receiver height.
\end{abstract}

\begin{IEEEkeywords}
5G, 3D topology, antenna radiation pattern, IoT, UAV,  uncoordinated network.
\end{IEEEkeywords}

\section{Introduction}
With the emerging of 5G wireless systems, network densification becomes crucial to improve data throughput \cite{gupta2015survey}. In the prospective networks, various types of Internet of things (IoT) devices, such as sensors, mobile phones, vehicles, are pervasively present and connected together. Massive access IoT is thereby the growing concept where ubiquitous devices communicate and interact with each other \cite{whitmore2015internet,ijaz2016enabling,al2015internet}. In order to satisfy the growing demand for IoT devices, various different technologies have been developed and standardized. In particular, 3GPP developed the narrowband IoT (NB-IoT) specifications~\cite{popli2018survey,rico2016overview}, while LoRa and Sigfox are introduced as alternative low-power wide-area network (LPWAN) technologies~\cite{sinha2017survey}.

% IG: This paragraph seems out of place; I removed, will check whether can integrate anything in later paragraphs.
% In order to support IoT network deployments, heterogeneous networks with  uncoordinated deployments is emerging \cite{merwaday2016handover,merwaday2014capacity}. Due to massive access of the IoT network,  interference management becomes critical \cite{teng2018resource, zucchetto2017uncoordinated}. Besides, due to unscheduled and distributed characteristic of the uncoordinated system as well as low-cost nature of IoT devices, many existing techniques that mitigate interference in the centralized networks are not available for IoT networks. For example, resource allocation, zero-forcing precoding, interference cancellation, interference alignment methods are hard to be applied in an uncoordinated network.

Future IoT deployments are expected to involve various different kinds of devices and applications. Among them, communication with aerial devices, such as cellular-connected unmanned aerial vehicles (UAVs), has recently received major attention~\cite{zeng2018cellular,xiao2016enabling}. UAVs have been considered as part of a 3D IoT network in~\cite{motlagh2017uav} for crowd surveillance purposes, while 3D scenarios for IoT deployments have been considered in \cite{wei2016gyro,ciftler2017iot} for RFID-based localization, and \cite{bujari2018standards} provides a broader overview with 3D IoT deployments. %To mitigate the severe interference in uncoordinated IoT networks, strategies that utilize carrier wave-form of the signal are commonly used. The concept of partially overlapping tone in filtered multi-tones (FMT) which allocates different frequency offsets in filter design is introduced in \cite{csahin2014partially,yilmaz2014game}. Another studied approach to suppress interference is based on sensing the channel and avoiding interference in a distributed fashion \cite{zheng2017adaptive}.
With massive deployments of IoT networks, interference management becomes a critical challenge~\cite{teng2018resource, zucchetto2017uncoordinated}, which especially has not been explored in detail in the 3D space.  

% Due to the unscheduled and distributed characteristic of the uncoordinated IoT networks, as well as low-cost nature of IoT devices, many existing techniques that mitigate interference in the centralized networks are not available for IoT networks. For example, resource allocation, zero-forcing precoding, interference cancellation, interference alignment methods are hard to be applied in an uncoordinated and low-cost IoT network.  In this paper,  considering both aerial and ground IoT nodes, we propose a new interference mitigation scheme in uncoordinated IoT networks that utilizes the 3D radiation pattern of dipole antenna. 

The effect of 3D radiation pattern  has been studied for massive MIMO beamforming in the literature, where angle dependent antenna gain incorporated with beamforming gain is studied~\cite{kammoun20153d,baianifar2017impact,li2013dynamic,rebato2019stochastic,3gpp}. 3D beamforming with UAVs have been explored in~\cite{geraci2018understanding,zhu20193}, which consider the effect of the 3D antenna radiation pattern combined with the 3D spatial beamforming. However, these approaches, and associated interference mitigation schemes such as zero-forcing precoding, interference cancellation, and interference alignment may not be suitable for tackling with interference problems in 3D space for low-cost and low-complexity IoT devices. Such IoT devices typically operate below $1$~GHz, and may employ only a single (or few) dipole antennas, and have limited computational capabilities. While dipole antenna radiation pattern with different configurations in the 3D space have been studied in~\cite{chen2018impact,shafi2006polarized,dao20113d,maeng2019interference}, there is no detailed analysis of interference characteristics and mitigation schemes with aerial equipment to our best knowledge.

In this paper,  considering both aerial and ground IoT nodes, we propose a new interference mitigation scheme in uncoordinated IoT networks that utilizes the 3D radiation pattern of dipole antenna. 
The main concept of our proposed interference mitigation scheme is that if we utilize the different antenna radiation pattern at the transmitter side depending on the 3D location of the receiver, we can suppress interference signal and enhance the desired signal. In general, in a 2D space topology we assume that the dipole antenna is aligned with $z$-axis, which generates omni-directional radiation pattern with respect to azimuth angle. However, this dipole setting, which may be common in typical low-cost IoT devices, cannot have onmi-directional radiation pattern in 3D topology  (as in networks including aerial nodes) due to power that varies with the elevation angle. On the other hand, by aligning dipole antenna with different direction, such as $y$-axis, we can obtain a different directivity of the radiation pattern.

%IG: I could not figure out how to smoothly connect this paragraph to the rest of the introduction. Does not fit into the flow. While referring to references, we should try compare/contrast with what is different in our work. I am not sure how tightly coupled the references in the papers below are with our work. Will remove for now, may include back more smoothly if we can get a major revision, or for future submission. 

% The distance based channel model with randomly located transmitters and receivers has been studied in the literature. In \cite{srinivasa2010distance}, the distribution function of the Euclidean distances in 2D space is derived. the 3D space distance based channel model also has been investigated in the literature. \cite{chandhar2017massive} suggests the distance model that the aerial receiver UAVs are randomly located in spherical shell with fixed location of a transmitter at $(0, 0, 0)$ Cartesian coordination. In \cite{chetlur2017downlink}, the paper studies the scenario that UAVs are located at aerial plane with fixed height and ground devices are located at the ground plane. The number of devices and the location of the devices are generated by random variables. In our work, we propose the distance based 3D topology channel design by uniform randomly generated azimuth angle and distance with fixed number of devices and the height.

In our previous work \cite{maeng2019interference}, we investigate the interference mitigation  scheme by the 3D radiation pattern in the 3D topology network with similar setting. In particular, we propose 2 and 3 dipole antenna schemes in order to generate various radiation patterns depending on antenna configurations, and show the  performance improvement by varying the height of the aerial devices and the proportion of the aerial devices by simulations. On the other hand, in this paper, different than~\cite{maeng2019interference}, we focus on both analytically and numerically showing the performance improvement based on the concrete 3D topology channel model and solving the ergodic achievable rate. The contributions of the present paper can be listed as follows.
\begin{itemize}
  \item[--] We propose a 3D topology channel model based on the antenna patterns and the location of the devices by  using uniformly random variable azimuth angle ($\phi$) and distance in the 2D plane ($r$). In addition, we derive the PDF of the elevation angle ($\theta$) and 3D distance ($R$) in order to calculate the distribution of the distance between transmitters and receivers.
  \item[--] Based on the PDF of the distance and angles, we derive the closed-form equations of the expectation of the channel gains, which incorporates pathloss, small-scale fading, and antenna gain.
  \item[--] We derive the ergodic achievable rate of the aerial receivers with different dipole antenna setting, and propose the cross-dipole antenna scheme that improves the rate compared with the conventional single dipole scheme.
  \item[--] By numerical results, we show that the proposed cross-dipole scheme outperforms the single dipole scheme, and the ergodic achievable rate grows as the height of the aerial receiver increases. In addition, we show the best ratio between the ground receiver and aerial receiver with the different height of the aerial receivers.
  \end{itemize}

The rest of this paper is organized as follows. Section~\ref{sec:system} provides the IoT system model with both ground and aerial IoT nodes. Section~\ref{sec:antenna effect} analyzes the effect of the 3D annenna patterns on the IoT link link qualities considering a multi-access communication scenario. Section~\ref{sec:scen2} derives the ergodic rates in such a multi-access IoT network, and proposes an antenna selection scheme for interference mitigation. Section~\ref{sec:results} provides numerical results for an IoT network with mixed ground/aerial links, and the last section concludes the paper.   

\section{System Model} \label{sec:system}

\begin{figure}[!t]
	\centering
	\vspace{-0.0in}
	\includegraphics[width=0.5\textwidth]{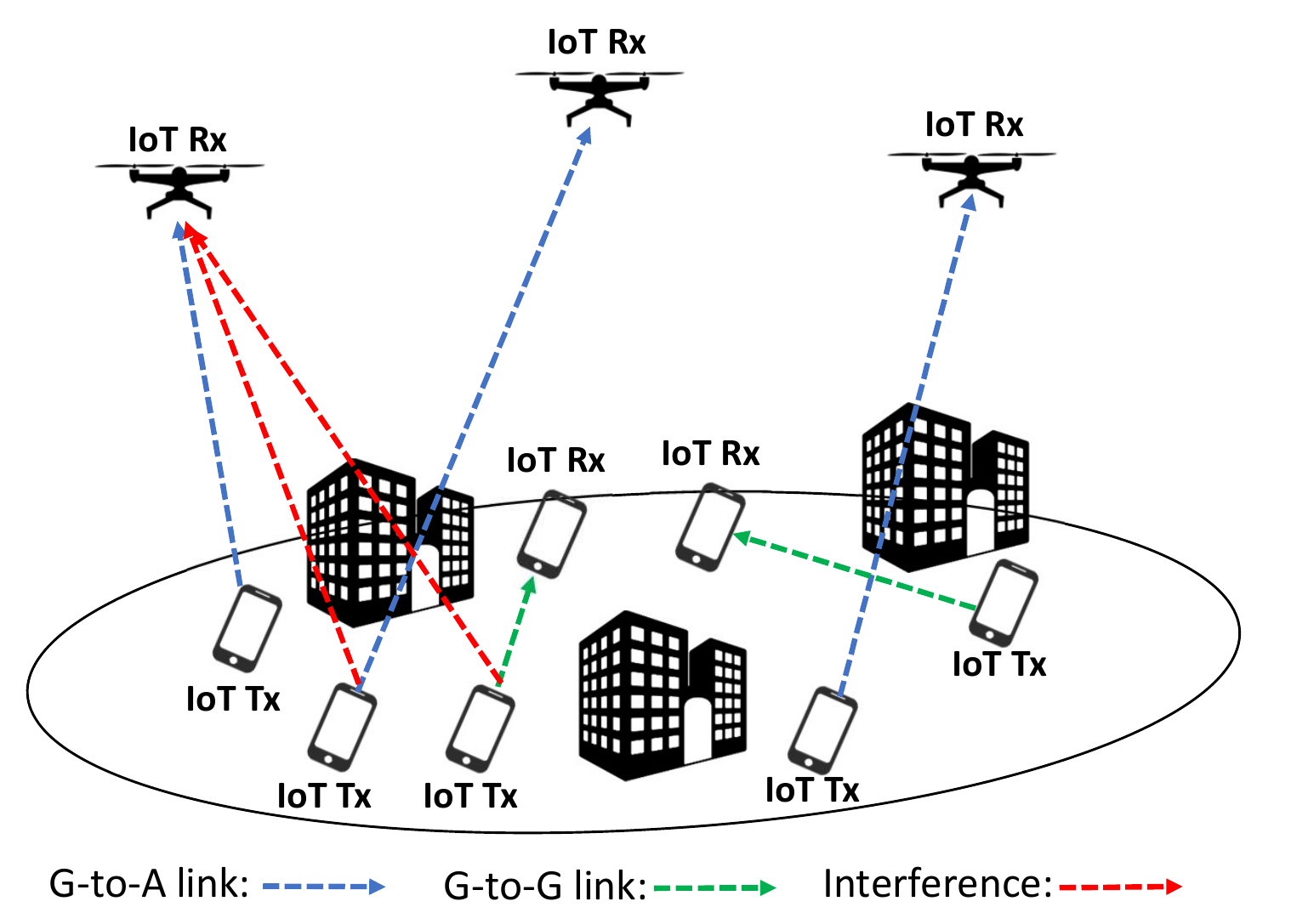}
	%\vspace{-0.15in}
	\caption{Illustration of the 3D topology of the IoT network with both ground and aerial nodes.}
	\label{fig:illu}
	%\vspace{-0.2in}
\end{figure}

In this study, we consider uncoordinated network with IoT devices as shown in Fig.~\ref{fig:illu}. Transmitter and receiver pairs are distributed in 3D space without resource management from a central base station. Thus, time and frequency resources are shared by transmitter/receiver (Tx/Rx) pairs. IoT devices are divided into ground and aerial devices depending on the typical altitude of the device. For example, the sensor is regarded as a ground device, and the UAV is considered as an aerial device. We consider the case that all IoT Txs are ground devices, while IoT Rxs are either ground or aerial devices. Thus, there are links between ground to ground (G-to-G), ground to air (G-to-A) in the networks.

\subsection{3D Topology Based Channel Model}
The channel between a transmitter and a receiver is modeled by pathloss, small-scale fading, and antenna gain. The distance between transmitter and receiver is easily derived by the 3D Cartesian coordination of two devices. Let $(x^{\rm Tx}_i,y^{\rm Tx}_i,z^{\rm Tx}_i)$, $(x^{\rm Rx}_i,y^{\rm Rx}_i,z^{\rm Rx}_i)$ denote the positions of Tx and Rx of $i_{\rm th}$ pair. Then, the distance between Tx and Rx is calculated as $d_{i,i}=\sqrt{(x^{\rm{Tx}}_i-x^{\rm{Rx}}_i)^2+(y^{\rm{Tx}}_i-y^{\rm{Rx}}_i)^2+(z^{\rm{Tx}}_i-z^{\rm{Rx}}_i)^2}$. Then, the free-space pathloss can be expressed as
\begin{align}\label{eq:pl}
\beta_{i,i}&= \left(\frac{\lambda}{4\pi d_{i,i}}\right)^2,
\end{align}
where $\lambda$ is the wave length of the signal. We denote $\alpha$ as the small scale fading coefficient, and $G^{\rm Tx}_i$, $G^{\rm Rx}_i$ are antenna gain of transmitter and receiver sides, respectively. The channel coefficient $g_{i,i}$ can be written by
\begin{align}\label{eq:chnl}
g_{i,i}&= \sqrt{PG_i^{\rm{Tx}}\beta_{i,i}G_i^{\rm{Rx}}}\alpha_{i,i},
\end{align}
where $P$ is signal power from transmitter. The magnitude of small scale fading coefficient follows Rayleigh distribution, $\alpha\sim \mathcal{CN}(0,1)$. Although we consider NLoS small scale fading model due to long-distance IoT link, air-to-ground channels can be also modeled by Rician fading that LoS component is dominant depending on the channel environment, as:
\begin{align}\label{eq:chnl_ri}
g_{i,i}&= \sqrt{PG_i^{\rm{Tx}}\beta_{i,i}G_i^{\rm{Rx}}}\left\{\sqrt{\frac{\kappa}{\kappa+1}}+\sqrt{\frac{1}{\kappa+1}}\alpha_{i,i}\right\},
\end{align}
where $\kappa$ is the power ratio between the LoS and NLoS components. We consider Rician fading channel model by simulation results in Section \ref{sec:rician}. The antenna gain $G_i^{\rm{Tx}}$, $G_i^{\rm{Rx}}$ is the function of the angle of departure (AoD) and the angle of arrival (AoA). The angle can be represented by the azimuth angle ($\phi$) and the elevation angle ($\theta$), and the function $G_i^{\rm{Tx}}$, $G_i^{\rm{Rx}}$ is changed depending on the antenna model.

\subsection{The Ergodic Achievable Rate}
The ergodic achievable rate of the aerial receiver can be represented by the channel gain from the connected transmitter, interference from other transmitters, and additive noise, as:
\begin{align}\label{eq:rate}
S_i&=\mathbb{E}\left\{\log_2\left(1+\frac{|g_{i,i}|^2}{\sum_{j\neq i}|g_{i,j}|^2+\sigma^2_n}\right)\right\},
\end{align}
where $S_i$ is the ergodic achievable rate of the aerial receiver ($i_{\rm th} $ receiver), and $\sigma^2_n$ is the noise variance. We can approximate the ergodic achievable rate in \eqref{eq:rate} as
\begin{align}\label{eq:rate_apx}
S_i&\overset{(a)}{\approx}\mathbb{E}\left\{\log_2\left(1+\frac{|g_{i,i}|^2}{\sum_{j\neq i}|g_{i,j}|^2}\right)\right\}\nonumber\\
&\overset{(b)}{\approx}\log_2\left(1+\frac{\mathbb{E}\{|g_{i,i}|^2\}}{\sum_{j\neq i}\mathbb{E}\{|g_{i,j}|^2\}}\right),
\end{align}
where $(a)$ comes from the assumption that the effect of noise is trivial in the interference dominant network, and $(b)$ comes from the approximation related with Jensen's inequality \cite{zhang2014power}, \cite{fan2015uplink}. Note that the approximation $(b)$ holds if $|g_{i,i}|^2$, $|g_{i,j}|^2$ are non-negative, and it is more accurate as the number of random variables increase.

\section{The Effect of Antenna Radiation Pattern on 3D Topology Networks} \label{sec:antenna effect}

\begin{figure}[!t]
	\centering
	\vspace{-0.0in}
	\includegraphics[width=0.5\textwidth]{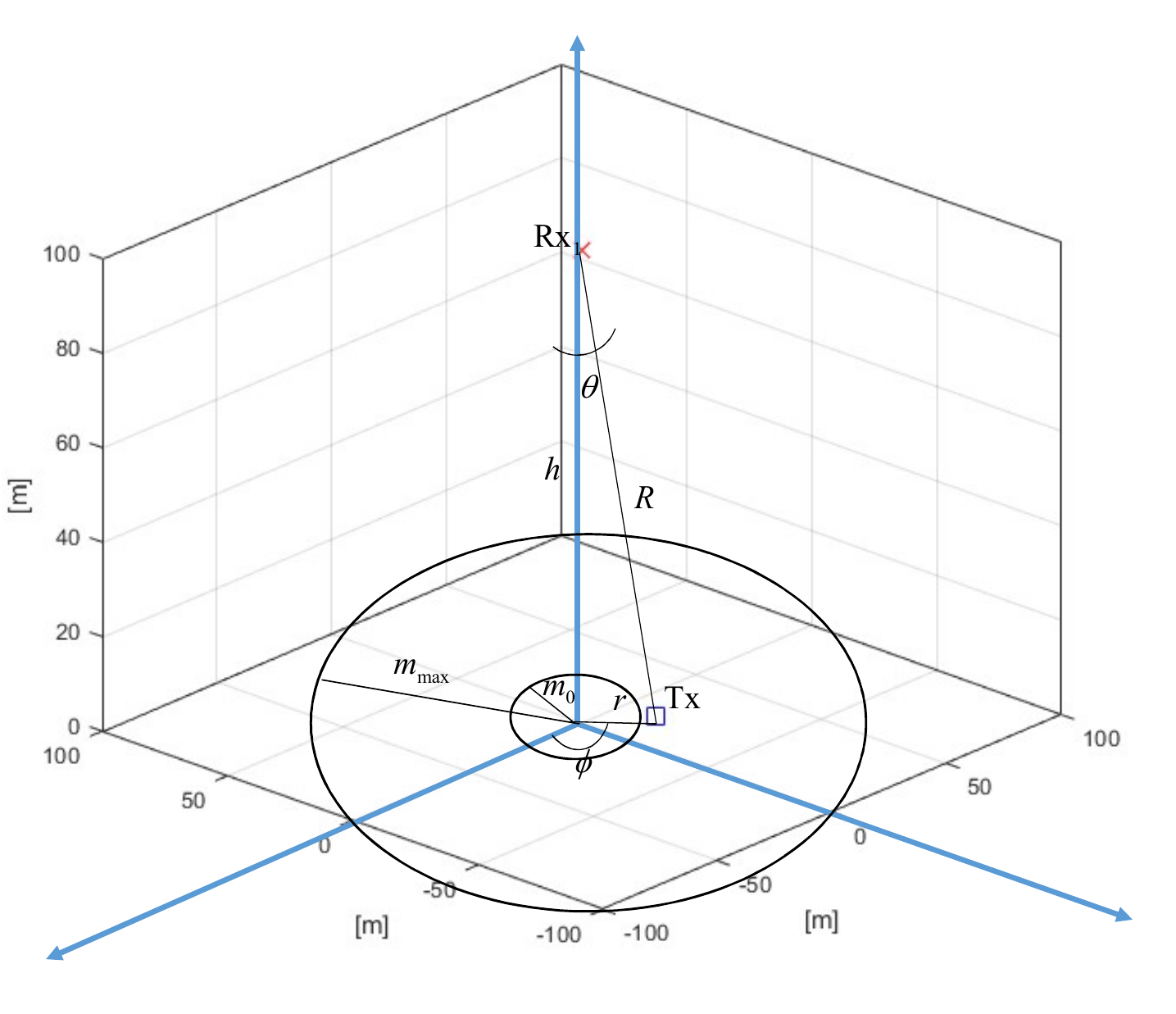}
	\vspace{-0.1in}
	\caption{The 3D topology of IoT network in stand-alone scenario.}
	\label{fig:top1}
	\vspace{0in}
\end{figure}

\begin{figure}[!t]
	\centering
	\vspace{-0.0in}
	\includegraphics[width=0.5\textwidth]{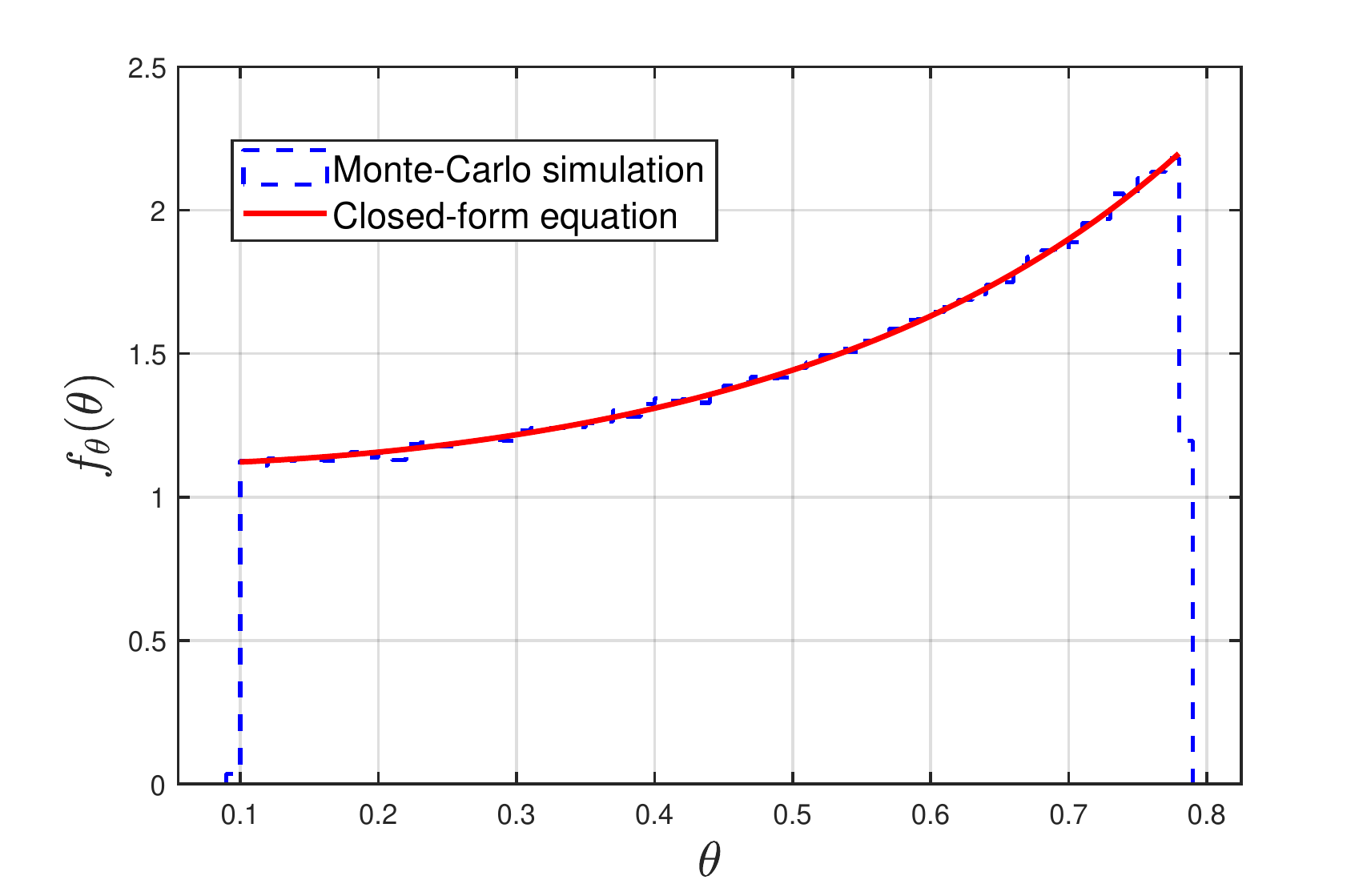}
	\vspace{-0.1in}
	\caption{The PDF of random variable $\theta$ on stand-alone scenario in \eqref{eq:pdf_th}. $m_0 = 10$, $m_{\rm max} = 100$, $h=100$.}
	\label{fig:pdf_th}
	%\vspace{-0.2in}
\end{figure}

In this section, we study the effect of 3D antenna radiation pattern in channel gain $|g_{i,i}|^2$ and achievable rate $S$. We assume that ground transmitters utilize cross-dipole antenna, and aerial receivers have onmi-directional radiation pattern antenna. The two dipole antennas are placed on $z$-axis and $y$-axis of Cartesian coordinate. Thus, the ground transmitters are able to transmit signal either $z$-axis dipole antenna or $y$-axis dipole antenna. At first, we consider stand-alone case, which means that only one Tx/Rx pair is located at a given space. Then, we expand it to multiple Tx/Rx scenario.

\subsection{Analysis of Stand-alone Scenario}

\textit{1) PDF of random variables related with the location of devices:} Let consider one Tx/Rx pair. The location of ground transmitter is decided by 2D circle radius ($r$) and azimuth angle ($\phi$), and the location of aerial receiver is fixed at $(0,0,h)$ Cartesian coordinate. The illustration of the 3D topology of the IoT network is shown in Fig.~\ref{fig:top1}. The random variables $r$, $\phi$ are independently uniformly distributed as
\begin{align}\label{eq:pdf rphi}
 f_r(r)&=\frac{1}{m_{\rm max}-m_0}, \;[m_0<r<m_{\rm max}],\nonumber\\
 f_{\phi}(\phi)&=\frac{1}{2\pi},\;[0<\phi<2\pi],
\end{align}
where $f_r(r)$, $f_{\phi}(\phi)$ are the  probability density function (PDF) of $r$ and $\phi$, while $m_0$,  $m_{\rm max}$ are the minimum and  the maximum radius of the circle, respectively. The elevation angle ($\theta$) can be represented as $\theta=\tan^{-1}(\frac{r}{h})$. Then, the PDF of random variable $\theta$ can be derived by
\begin{align}\label{eq:pdf_th}
 \rm{d}\theta&=\frac{h}{r^2+h^2}{\rm d}r,\nonumber\\
 f_{\theta}(\theta)&\overset{(a)}{=}\sum f_{r}(h\tan\theta)\left|\frac{{\rm d}r}{\rm d\theta}\right|\nonumber\\
 &=\sum f_{r}(h\tan\theta)\left|\frac{h^2\tan^2\theta+h^2}{h}\right|\nonumber\\
 &=\sum f_{r}(h\tan\theta)(h\tan^2\theta+h)\nonumber\\
 &=\frac{h}{(m_{\rm max}-m_0)}\tan^2\theta+\frac{h}{(m_{\rm max}-m_0)},\nonumber\\ &\left[\tan^{-1}\left(\frac{m_0}{h}\right)<\theta<\tan^{-1}\left(\frac{m_{\rm max}}{h}\right)\right],
\end{align}
where $(a)$ comes from the PDF transformation function. The distance between Tx and Rx is easily obtained as $R=\frac{h}{\cos\theta}$. The PDF of $\theta$ in \eqref{eq:pdf_th} is shown in Fig.~\ref{fig:pdf_th}, which is confirmed by Monte-Carlo simulation.

\begin{figure}[!t]
	\centering
		\vspace{-0.0in}
	\includegraphics[width=0.5\textwidth]{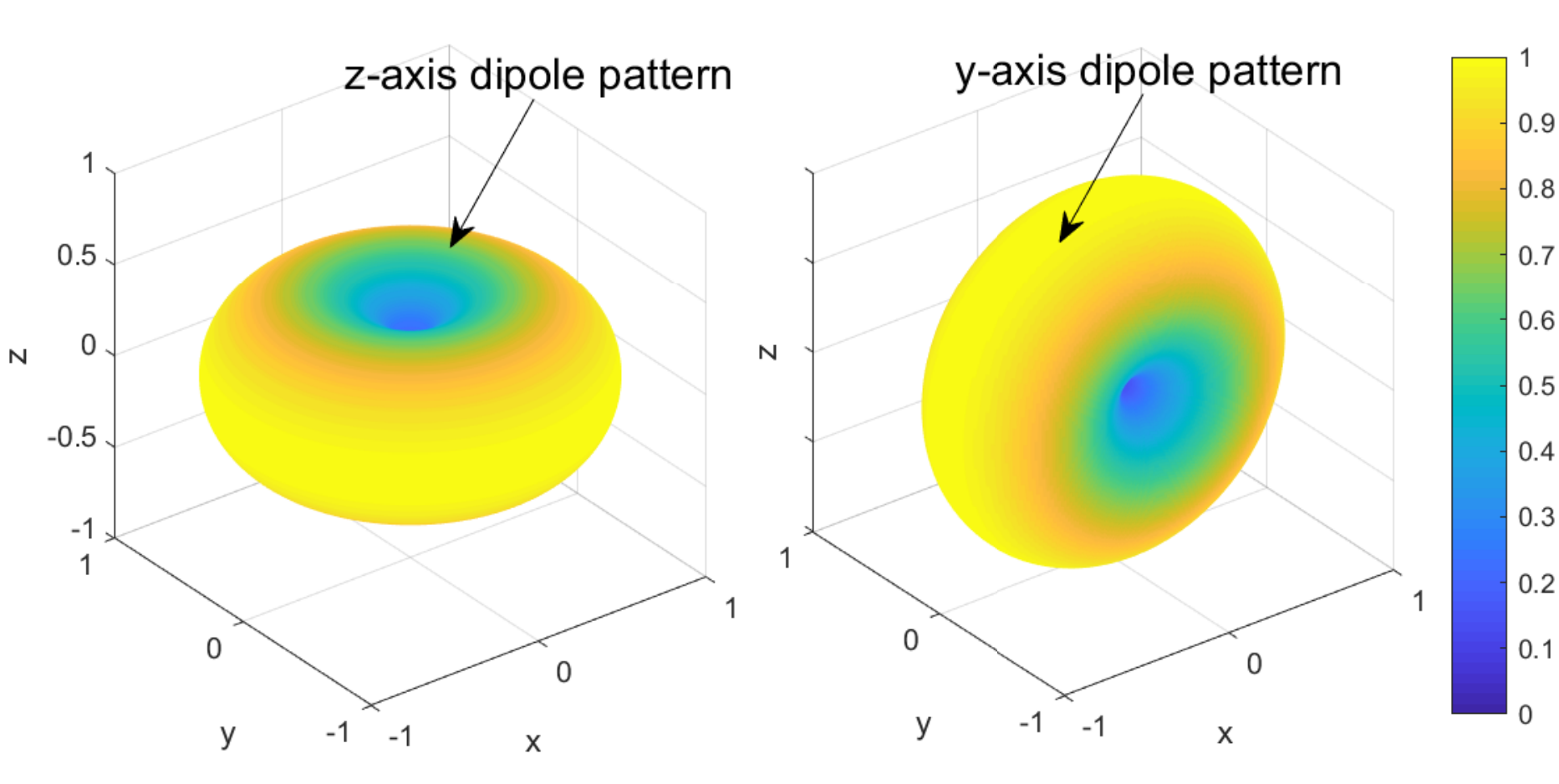}
	\vspace{-0.02in}
	\caption{The field pattern of dipole antenna on Cartesian coordinate (normalized magnitude).}
	\label{fig:pattern}
	\vspace{-0.0in}
\end{figure}

\textit{2) Antenna radiation pattern of cross-dipole antenna:} The radiation pattern of dipole antenna is interpreted by normalized antenna field pattern $F$. If we place the dipole antenna on $z$-axis, the radiation pattern is onmi-directional to azimuth angle ($\phi$). The normalized antenna field pattern of $z$-axis dipole antenna is written as \cite{balanis2016antenna}, \cite{chandhar2017massive}:
\begin{align}\label{eq:F_z}
F_{\rm{z}}(\theta)&=\frac{\cos\left(\frac{\pi f_0 d_{\rm{len}}}{c}\cos\theta\right)-\cos\left(\frac{\pi f_0d_{\rm{len}}}{c}\right)}{\sin\theta},
\end{align}
where $d_{\rm{len}}$, $c$, $f_0$ denote the length of dipole antenna, the speed of light, and carrier frequency, respectively. If we assume half-wave length dipole antenna ($d_{\rm{len}}=\frac{\lambda}{2}$), $\frac{\pi f_0 d_{\rm{len}}}{c}=\frac{\pi}{2}$ holds. Then, \eqref{eq:F_z} can be rewritten as
\begin{align}\label{eq:F_z re}
F_{\rm{z}}(\theta)&=\frac{\cos\left(\frac{\pi}{2}\cos\theta\right)}{\sin\theta}.
\end{align}
Let dipole antenna be placed on $y$-axis. The angle between the dipole antenna direction and the signal propagation direction is $\cos^{-1}(\hat{\textbf{s}}\cdot\hat{\textbf{y}})=\cos^{-1}(\sin(\theta)\sin(\phi))$, where $\hat{\textbf{s}}$, $\hat{\textbf{y}}$ are the unit vector of the signal and $y$-axis. Then, the normalized antenna field pattern of $y$-axis dipole antenna is given by
\begin{align}\label{eq:F_y}
F_{\rm{y}}(\theta,\phi)&=\frac{\cos\left(\frac{\pi}{2}\cos\left(\cos^{-1}(\sin(\theta)\sin(\phi))\right)\right)}{\sin\left(\cos^{-1}(\sin(\theta)\sin(\phi))\right)}.
\end{align}
Note that the antenna field pattern of $y$-axis is the function of both the azimuth angle ($\phi$) and the elevation angle ($\theta$). It means that the antenna gain can be changed by varying the azimuth angle as well as the elevation angle. The field patterns of $z$-axis, $y$-axis dipole antenna from \eqref{eq:F_z re}, \eqref{eq:F_y} on Cartesian coordinate are shown in Fig.~\ref{fig:pattern}.

\textit{3) The expectation of the channel gain:} The statistical knowledge of the channel gain is important in order to obtain the achievable rate of the user or system. We hence derive the expectation value of the channel gain $\mathbb{E}\{|g_{i,i}|^2\}$ here, to calculate the ergodic achievable rate later on. If we consider $z$-axis dipole antenna on ground transmitter and omni-directional antenna on aerial receiver, we can express the expectation of the channel gain from \eqref{eq:pl}, \eqref{eq:chnl} as 
\begin{align}\label{eq:exp_Fz}
\mathbb{E}\{|g_{i,i}|^2\}_{\rm z}&=\mathbb{E}\{ PF_{\rm z}^2\beta\alpha^2\}\nonumber\\
&\overset{(a)}{=}\frac{P\lambda^2}{16\pi^2 }\mathbb{E}\left\{\alpha^2\right\}\mathbb{E}\left\{\frac{(F_{\rm{z}})^2}{R^2}\right\}\nonumber\\
&\overset{(b)}{=}k_1\mathbb{E}\left\{\frac{(F_{\rm{z}})^2}{R^2}\right\}.
\end{align}
Since $\alpha$ is independent of the positions of devices, $(a)$ holds, and $(b)$ comes from $k_1=\frac{P\lambda^2}{16\pi^2 }$,  $\mathbb{E}\left\{\alpha^2\right\}=1$.
By using \eqref{eq:F_z re}, \eqref{eq:exp_Fz} can be rewritten as
\begin{align}\label{eq:EgFz}
&\mathbb{E}\{|g_{i,i}|^2\}_{\rm z}\nonumber\\
&=k_1\mathbb{E}\left\{\left(\frac{\cos\left(\frac{\pi}{2}\cos\theta\right)}{\sin\theta}\right)^2\frac{\cos^2\theta}{h^2}\right\}\nonumber\\
&\overset{(a)}{=}k_1\int\left(\frac{\cos\left(\frac{\pi}{2}\cos\theta\right)}{h\sin\theta}\right)^2\cos^2\theta\ f_\theta(\theta){\rm d}\theta\nonumber\\
&\overset{(b)}{=}\frac{k_1}{(m_{\rm max}-m_0)h}\int^{\tan^{-1}\left(\frac{m_{\rm max}}{h}\right)}_{\tan^{-1}\left(\frac{m_0}{h}\right)}\left(\frac{\cos\left(\frac{\pi}{2}\cos\theta\right)}{\sin\theta}\right)^2\nonumber\\
&\qquad\times\cos^2\theta\ (1+\tan^2(\theta)){\rm d}\theta\nonumber\\
&\overset{(c)}{=}\frac{k_1}{(m_{\rm max}-m_0)h}\int^{\tan^{-1}\left(\frac{m_{\rm max}}{h}\right)}_{\tan^{-1}\left(\frac{m_0}{h}\right)}\left(\frac{\cos\left(\frac{\pi}{2}\cos\theta\right)}{\sin\theta}\right)^2{\rm d}\theta,
\end{align}
where $(a)$ comes from the definition of the expectation, $(b)$ comes from \eqref{eq:pdf_th}, and $(c)$ comes from $1+\tan^2(\theta)=\sec^2(\theta)$. Since there is no closed-form equation for \eqref{eq:EgFz}, we utilize asymptotic behavior to approximate the equation. If the height of the aerial receiver ($h$) goes to infinity, the interval of  the integral with respect to $\theta$ go to $0$; $\tan^{-1}(\frac{m_{\rm max}}{h})\to0, \text{ as } h\to\infty$. Then, we can apply Taylor series approximation at $\theta=0$,

\begin{align}\label{eq:apx_Fz}
&\mathbb{E}\{|g_{i,i}|^2\}_{\rm z}\nonumber\\
&\approx\frac{k_1}{(m_{\rm max}-m_0)h}\int^{\tan^{-1}\left(\frac{m_{\rm max}}{h}\right)}_{\tan^{-1}\left(\frac{m_0}{h}\right)}\left(\frac{\pi^2\theta^2}{16}\right){\rm d}\theta,\nonumber\\
&=\frac{\pi^2k_1\left[\left\{\tan^{-1}\left(\frac{m_{\rm max}}{h}\right)\right\}^3-\left\{\tan^{-1}\left(\frac{m_0}{h}\right)\right\}^3\right]}{48(m_{\rm max}-m_0)h},
\end{align}
where $\left(\frac{\pi^2\theta^2}{16}\right)$ is the first term of Taylor series.

\begin{figure}[!t]
	\centering
	\vspace{-0.0in}
	\includegraphics[width=0.5\textwidth]{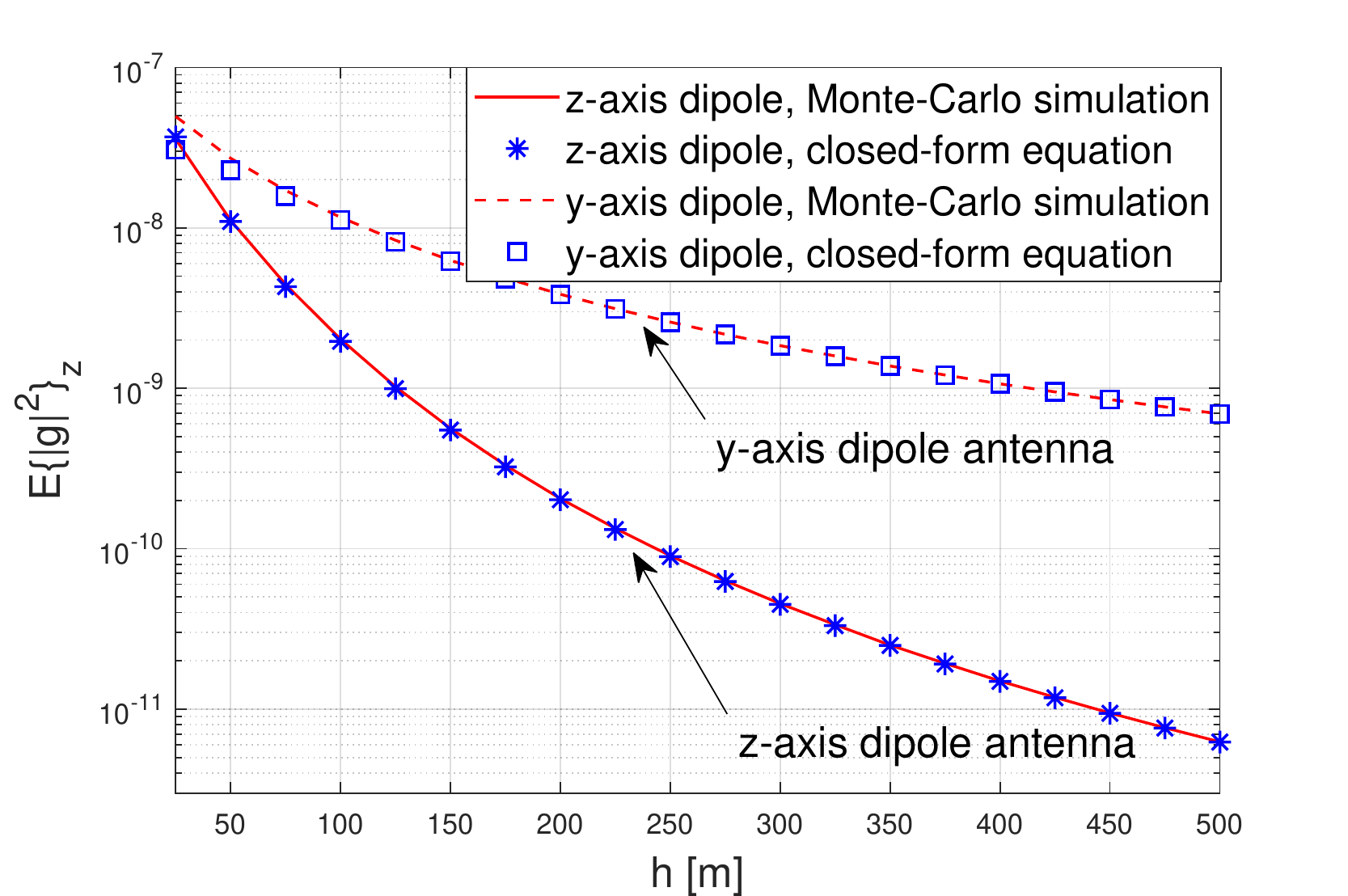}
	\vspace{-0.02in}
	\caption{The expectation of the channel gain verse height ($h$) with different dipole antenna placement on stand-alone scenario in \eqref{eq:apx_Fz}, \eqref{eq:apx_Fy}, and $m_0 = 10$, $m_{\rm max} = 100$.}
	\label{fig:chnl_sa}
	%\vspace{-0.2in}
\end{figure}

Now, if we consider $y$-axis dipole antenna for the ground transmitter, we can derive the channel gain by similar way.
\begin{align}\label{eq:EgFy}
&\mathbb{E}\{|g_{i,i}|^2\}_{\rm y}\nonumber\\
&=k_1\mathbb{E}\left\{\frac{(F_{\rm{y}})^2}{R^2}\right\}\nonumber\\
&\overset{(a)}{=}k_1\int\int\left(\frac{\cos\left(\frac{\pi}{2}\cos(\cos^{-1}(\sin(\theta)\sin(\phi)))\right)}{h\sin\left(\cos^{-1}(\sin(\theta)\sin(\phi))\right)}\right)^2\nonumber\nonumber\\
&\qquad\times\cos^2\theta\ f_\theta(\theta)f_\phi(\phi){\rm d}\theta {\rm d}\phi\nonumber\\
&\overset{(b)}{=}\frac{k_1}{2\pi h(m_{\rm max}-m_0)}\int^{2\pi}_0\int^{\tan^{-1}\left(\frac{m_{\rm max}}{h}\right)}_{\tan^{-1}\left(\frac{m_0}{h}\right)}\nonumber\\
&\qquad\times \left(\frac{\cos\left(\frac{\pi}{2}\cos(\cos^{-1}(\sin(\theta)\sin(\phi)))\right)}{\sin\left(\cos^{-1}(\sin(\theta)\sin(\phi))\right)}\right)^2{\rm d}\theta {\rm d}\phi,
\end{align}
where $(a)$ comes from the definition of the expectation and \eqref{eq:F_y}, and $(b)$ comes from the PDF functions in \eqref{eq:pdf rphi}. In the similar manner, since there is no closed-form equation for \eqref{eq:EgFy}, if the height ($h$) goes to infinity, we can apply Taylor series approximation at $\theta=0$, to obtain
\begin{align}\label{eq:apx_Fy}
&\mathbb{E}\{|g_{i,i}|^2\}_{\rm y}\nonumber\\
&\approx\frac{k_1}{2\pi h(m_{\rm max}-m_0)}\int^{2\pi}_0\int^{\tan^{-1}\left(\frac{m_{\rm max}}{h}\right)}_{\tan^{-1}\left(\frac{m_0}{h}\right)}\nonumber\\
&\qquad\times \left(1-\frac{1}{4}(\pi^2-4)\sin^2(\phi)\theta^2\right){\rm {\rm d}}\theta d\phi\nonumber\\
&=\frac{k_1}{2\pi h(m_{\rm max}-m_0)}\int^{2\pi}_{0}\nonumber\\
&\times\left(\frac{(4-\pi^2)\sin^2(\phi)\left[\{\tan^{-1}\left(\frac{m_{\rm max}}{h}\right)\}^3-\{\tan^{-1}\left(\frac{m_0}{h}\right)\}^3\right]}{12}\right.\nonumber\\
&\qquad\left.+\tan^{-1}\left(\frac{m_{\rm max}}{h}\right)-\tan^{-1}\left(\frac{m_0}{h}\right)\right){\rm d}\phi\nonumber\\
&=\frac{k_1}{2\pi h(m_{\rm max}-m_0)}\nonumber\\
&\qquad\times\left(\frac{(4\pi-\pi^3)\left[\{\tan^{-1}\left(\frac{m_{\rm max}}{h}\right)\}^3-\{\tan^{-1}\left(\frac{m_0}{h}\right)\}^3\right]}{12}\right.\nonumber\\
&\qquad\qquad\left.+2\pi\tan^{-1}\left(\frac{m_{\rm max}}{h}\right)-2\pi\tan^{-1}\left(\frac{m_0}{h}\right)\right),
\end{align}
where $\left(1-\frac{1}{4}(\pi^2-4)\sin^2(\phi)\theta^2\right)$ is the first and the second terms of Taylor series at $\theta=0$ with fixed $\phi$. The tightness of the approximated close-form equations \eqref{eq:apx_Fz}, \eqref{eq:apx_Fy} is shown by simulation in Fig.~\ref{fig:chnl_sa}. It is observed that \eqref{eq:apx_Fz} is really close to the exact value even if the height ($h$) is low, and \eqref{eq:apx_Fy} becomes close to the exact value as $h$ increases. Note that the approximations should be more accurate at higher height, since we use $h\to\infty$.

\subsection{Analysis of Multiple Tx Scenario}\label{sec:scen1}
Let us consider multiple transmitters on the ground, and an aerial receiver is fixed at $(0, 0 , h)$ Cartesian coordinate. Let us assume that one ground transmitter which is connected with the aerial receiver utilizes either $z$-axis dipole antenna or $y$-axis dipole antenna, and all other ground transmitters select $z$-axis dipole antenna. We compare the achievable rate $S$ of the aerial receiver depending on the dipole antenna selection from the connected ground transmitter. As mentioned before, ground transmitters have cross-dipole antenna and they are capable of selecting either $z$-axis or $y$-axis dipole antenna. Intuitively, we can observe from the antenna field patterns that $z$-axis dipole achieves higher antenna gain to ground receivers, and $y$-axis dipole antenna obtains better performance to aerial receivers.

Let consider the case that the ground transmitter connected with the aerial receiever utilizes z-dipole antenna. Then, $\mathbb{E}\{|g_{i,i}|^2\}$ is equal to $\mathbb{E}\{|g_{i,i}|^2\}_{\rm z}$ in \eqref{eq:apx_Fz}. Since other transmitters utilize z-dipole antenna, $\mathbb{E}\{|g_{i,j}|^2\}$ is also equal to $\mathbb{E}\{|g_{i,i}|^2\}_{\rm z}$. If we assume that the number of pairs of Tx/Rx is $K$ and substitute $\mathbb{E}\{|g_{i,i}|^2\}_{\rm z}$ in \eqref{eq:apx_Fz} for $\zeta_{\rm z}$, the ergodic achievable rate of the aerial receiver from \eqref{eq:rate_apx} can be written as
\begin{align}\label{eq:rate_Fz}
\{S_i\}_{\rm z}&\approx\log_2\left(1+\frac{\zeta_z}{(K-1)\zeta_z}\right)=\log_2\left(1+\frac{1}{(K-1)}\right).
\end{align}
Note that the ergodic achievable rate with z-dipole antenna is only depending on the number of pairs ($K$) in the network. The reason for this is that either desired signal or interference signal has the same statistical values. 

\begin{figure}[!t]
	\centering
	\vspace{-0.0in}
	\includegraphics[width=0.5\textwidth]{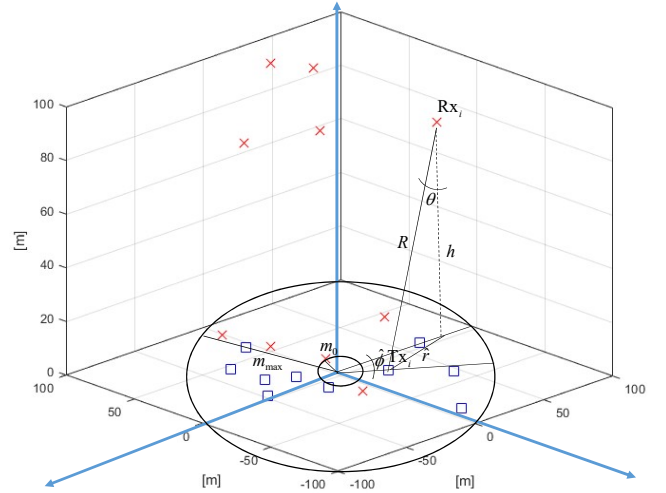}

	\caption{The 3D topology  design in IoT network on multiple Tx/Rx pairs scenario.}
	\label{fig:top2}
	%\vspace{-0.3in}
\end{figure}

\begin{figure*}[!t]
	\centering
	\hspace{-0.4in}
	\subfloat[The PDF of $\hat{\phi}$]{\includegraphics[width=0.38\textwidth]{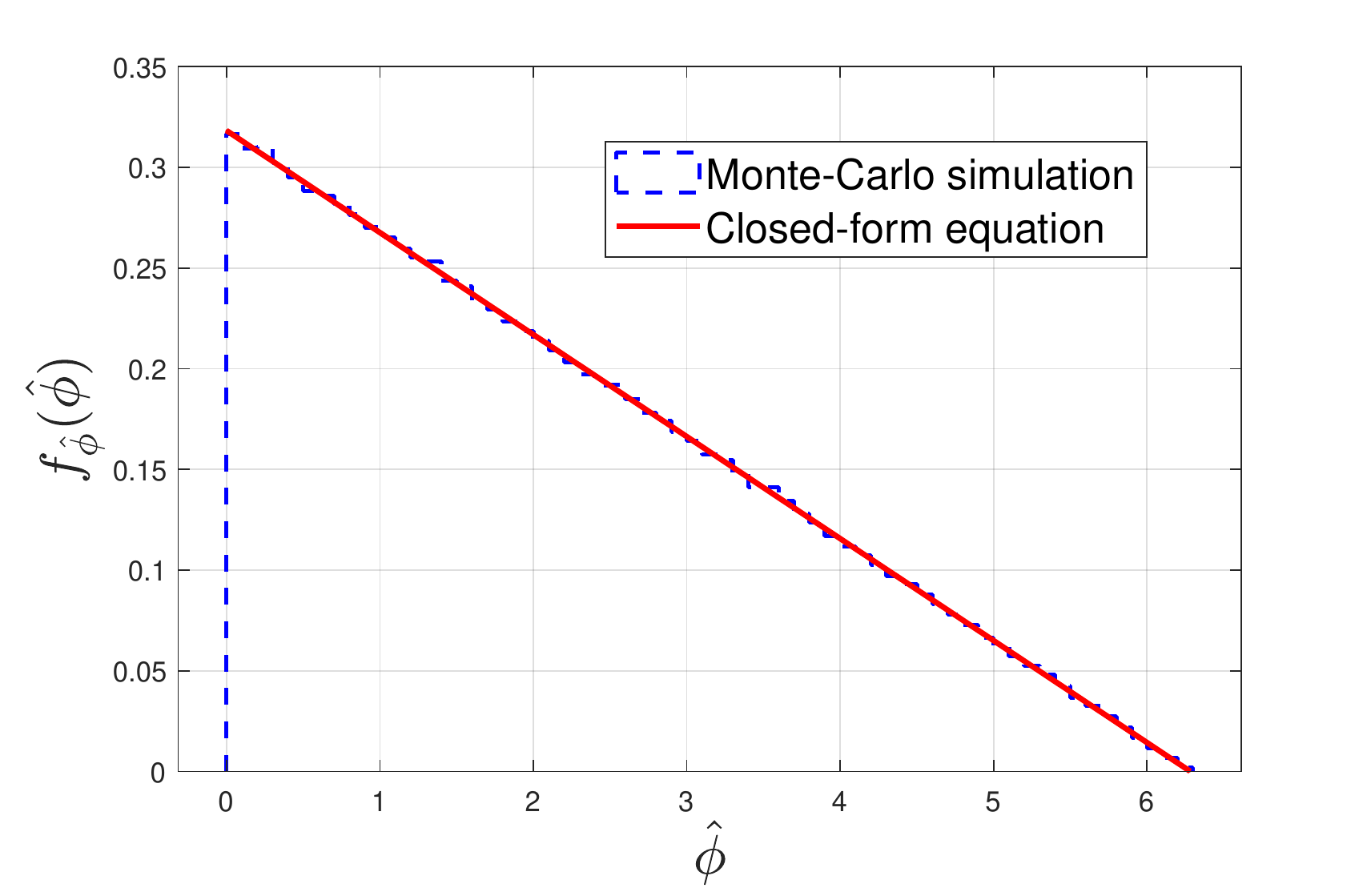}}
	\hspace{-0.40in}
	\subfloat[The PDF of $\hat{r}$]{\includegraphics[width=0.38\textwidth]{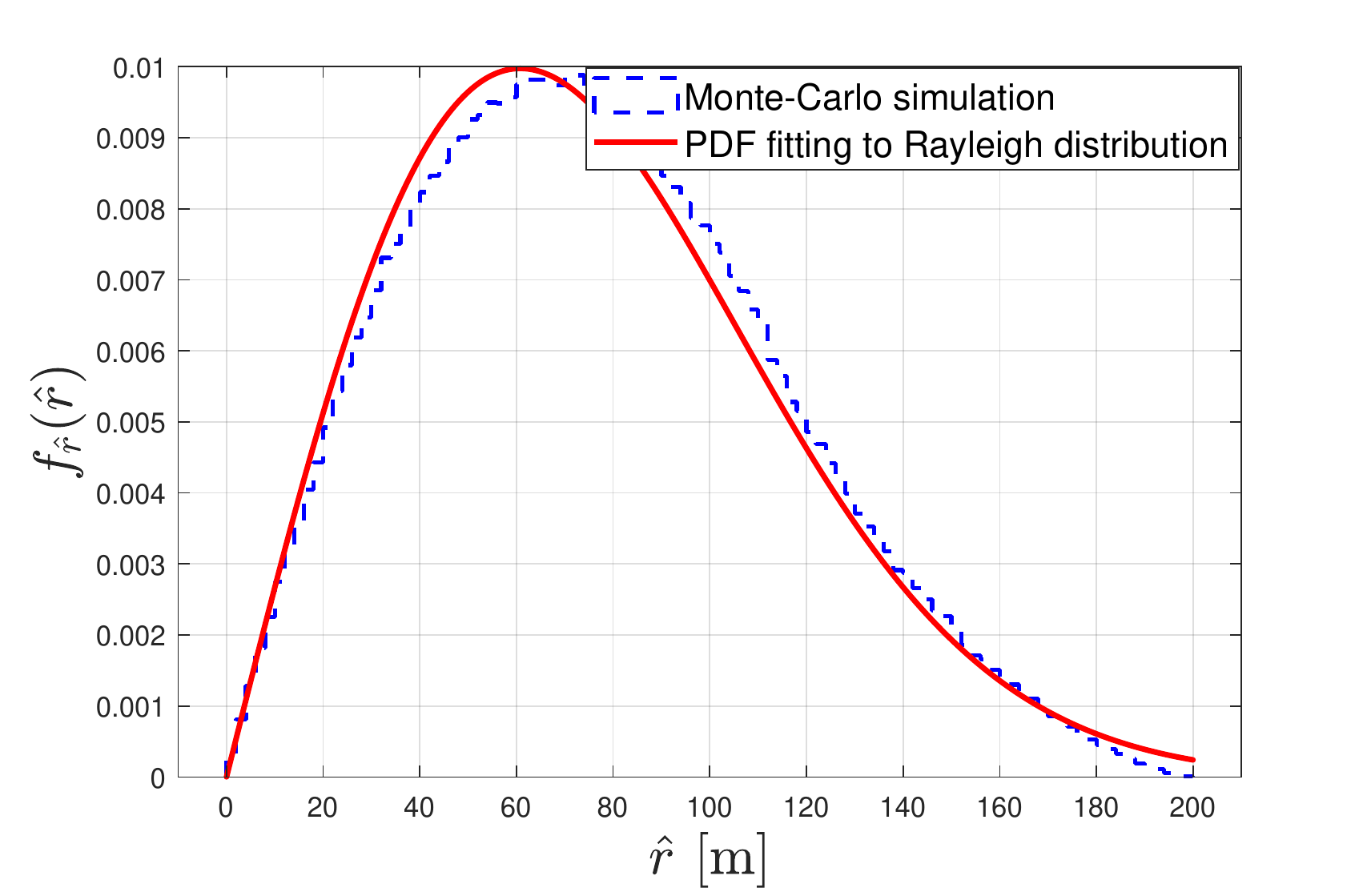}\label{fig:pdf_2_r}}
	\hspace{-0.30in}
	\subfloat[The PDF of $\theta$]{\includegraphics[width=0.38\textwidth]{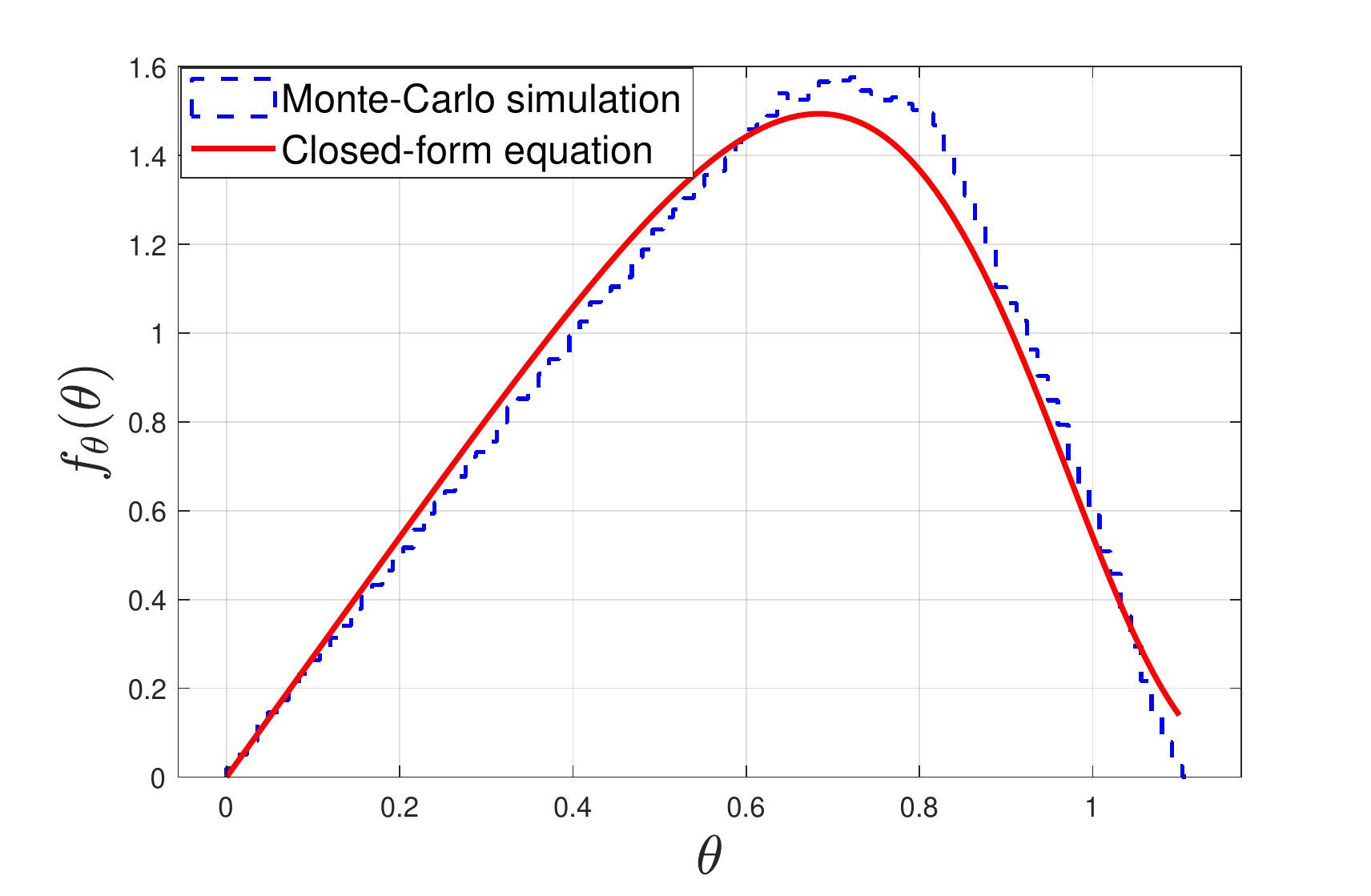}}
	\hspace{-0.25in}
	\caption{The PDF of random variables on multiple Tx/Rx scenario in \eqref{eq:pdf_phi2}, \eqref{eq:pdf_r2}, \eqref{eq:pdf_th2}, where $m_0 = 10$, $m_{\rm max} = 100$, $h=100$.}
	\label{fig:pdf_2}
	\vspace{-0.0in}
\end{figure*}

Next, let us consider the case that the ground transmitter utilizes y-dipole antenna. At this time, the channel gain from the desired signal  $\mathbb{E}\{|g_{i,i}|^2\}$ is equal to $\mathbb{E}\{|g|^2\}_{\rm y}$ in \eqref{eq:apx_Fy}. Then, after we substitute $\mathbb{E}\{|g|^2\}_{\rm y}$ for $\zeta_{\rm y}$, the ergodic achievable rate of the aerial receiver can be expressed as
\begin{align}\label{eq:rate_Fy}
\{S_i\}_{\rm y}&\approx\log_2\left(1+\frac{\zeta_{\rm y}(h)}{(K-1)\zeta_{\rm z}(h)}\right),
\end{align}
where channel gains $\zeta_{\rm y}(h), \zeta_{\rm z}(h)$ are the function of the height ($h$). We interestingly observe that the ergodic achievable rate from \eqref{eq:rate_Fy} improves as the height increases, while the ergodic achievable rate from \eqref{eq:rate_Fz} keeps constant as a function of the height, which is shown by simulations in Section \ref{sec:results}. It means that we can improve the performance of the achievable rate by utilizing different antenna radiation pattern depending on the type of the devices. For example, we can set the ground transmitters which are connected with the ground receiver to choose $z$-axis dipole antenna, and transmitters which are connected with the aerial receiver to select $y$-axis dipole antenna. Besides, we can obtain better performance, as the height of the aerial receiver is higher. 

\section{Analysis on Multiple Tx/Rx Pairs Scenario}\label{sec:scen2}
In this section, we extend the scenario to multiple Tx/Rx pairs without the limitation of the receivers' positions. It means that the receivers are not fixed at $(0,0,h)$ Cartesian coordinate, but decided by random variables. Moreover, receivers could be either ground receiver or aerial receiver. We assume that the height of all aerial receivers is fixed to $h$. Fig.~\ref{fig:top2} shows the new illustration of 3D topology design in IoT network with multiple Tx/Rx pairs scenario. 

\subsection{PDF of Random Variables Related with the Location of Devices}
At first, the location of the individual transmitters and receivers are decided by independently generated random variables. The PDF of uniformly distributed random variables that indicates the position of the devices are given by
\begin{align}\label{eq:pdf rphi m}
 f_{r^{\rm Tx}}(r^{\rm Tx})&=\frac{1}{m_{\rm max}-m_0}, \;[m_0<r^{\rm Tx}<m_{\rm max}],\nonumber\\
  f_{r^{\rm Rx}}(r^{\rm Rx})&=\frac{1}{m_{\rm max}-m_0}, \;[m_0<r^{\rm Rx}<m_{\rm max}],\nonumber\\
 f_{\phi^{\rm Tx}}(\phi^{\rm Tx})&=\frac{1}{2\pi},\;[0<\phi^{\rm Tx}<2\pi],\nonumber\\
  f_{\phi^{\rm Rx}}(\phi^{\rm Rx})&=\frac{1}{2\pi},\;[0<\phi^{\rm Rx}<2\pi],
\end{align}
where $r^{\rm Tx}$, $r^{\rm Rx}$ are the 2D circle radius of transmitters and receivers, and $\phi^{\rm Tx}$, $\phi^{\rm Rx}$ are azimuth angle of transmitters and receivers. Since the location of both transmitters and receivers are determined by random variables, we need to calculate the relative difference of the Tx/Rx random variables in order to derive the distance between transmitters and receivers ($R$):
\begin{align}\label{}
\hat{\phi}&=|\phi^{\rm Rx}-\phi^{\rm Tx}|,\nonumber\\
\hat{r}&=\sqrt{(r^{\rm Tx})^2+(r^{\rm Rx})^2-2r^{\rm Tx}r^{\rm Rx}\cos\hat{\phi}},
\end{align}
where $\hat{\phi}$ is relative difference of azimuth angle between Tx and Rx, $\hat{r}$ is distance between Tx and Rx at 2D ground plane. The equation of $\hat{r}$ comes from the law of cosines in trigonometry. The PDF of $\hat{\phi}$ can be expressed as
\begin{align}\label{eq:pdf_phi2}
f_{\hat{\phi}}(\hat{\phi})&=\frac{1}{2\pi^2}(2\pi-\hat{\phi}), \;[0<\hat{\phi}<2\pi],
\end{align}
where \eqref{eq:pdf_phi2} is derived from the property that the PDF of the sum of the two independent random variables is the convolution of the PDF of two random variables. The closed-form equation of the PDF of $\hat{r}$ is hard to obtain. Alternatively, we obtain the approximated form of the PDF by using the PDF fitting tool in MATALAB. By simulation results, we observe that the PDF of $\hat{r}$ fits closely to the Rayleigh distribution:
\begin{align}\label{eq:pdf_r2}
f_{\hat{r}}(\hat{r})&\approx\frac{\hat{r}}{b^2}e^{\frac{-\hat{r}^2}{2b^2}},
\end{align}
where the coefficient $b=60.7994$ in the fitted PDF, when $m_0=10$, $m_{\rm max}=100$. Then, we can derive the PDF of $\theta$ from the equation $\hat{r}=h\tan(\theta)$ as follows:
\begin{align}\label{eq:pdf_th2}
f_{\theta}(\theta)&=\sum f_{\hat{r}}(h\tan\theta)\left|\frac{d\hat{r}}{d\theta}\right|\nonumber\\
 &=\sum f_{\hat{r}}(h\tan\theta)\left|\frac{h^2\tan^2\theta+h^2}{h}\right|\nonumber\\
 &=\frac{h^2\tan\theta}{b^2}e^{-\frac{h^2\tan^2\theta}{2b^2}}\sec^2\theta,\;[0<\theta<\tan^{-1}(\frac{2m_{\rm max}}{h})]~.
 \end{align}
 
\vspace{-0.07in}
\noindent The closed-form equations of PDF of $\hat{\phi}$, $\hat{r}$, $\theta$ are simulated in Fig.~\ref{fig:pdf_2}, which are compared with Monte-Carlo simulations. In Fig.~\ref{fig:pdf_2}\subref{fig:pdf_2_r}, it is observed that fitted PDF of $\hat{r}$ to Rayleigh distribution (red solid line) closely matches to the exact distribution (blue histogram). 
\vspace{-0.0in}
\subsection{The Expectation of the Channel Gain}
The expectation of channel gains can be calculated by the new PDF of random variables. By the similar analysis to the stand-alone scenario, 
we can obtain the expectation of the channel gain in case of both the ground transmitter with $z$-axis dipole antenna and the ground transmitter with $y$-axis dipole antenna, as follows:
\begin{align}\label{eq:chnl_Fz2}
&\mathbb{E}\{|g_{i,i}|^2\}_{\rm z}\nonumber\\
&\approx\frac{-3\pi^2k_1e^{-(3k_2)/8}}{512b^2k_2}\left(\sqrt{6\pi k_2}\text{erfi}\left(\frac{\sqrt{k_2}(4\theta^2+3)}{2\sqrt{6}}\right)\right.\nonumber\\
&\left.\left.-4e^{\frac{1}{24}k_2(4\theta^2+3)^2}\right)\right|^{\tan^{-1}\left(\frac{2m}{h}\right)}_{0}=\zeta_{\rm z},
\end{align}
\begin{align}\label{eq:chnl_Fy2}
&\mathbb{E}\{|g_{i,i}|^2\}_{\rm y}\nonumber\\
&\approx\left(\frac{-3k_1e^{-(3k_2)/8}}{64\pi^2b^2k_2}\right)\left(\frac{5\pi^2}{3}-\frac{\pi^4}{4}\right)\left(\sqrt{6\pi k_2}\right.\nonumber\\
&\times\left.\left.\text{erfi}\left(\frac{\sqrt{k_2}(4\theta^2+3)}{2\sqrt{6}}\right)-4e^{\frac{1}{24}k_2(4\theta^2+3)^2}\right)\right|^{\tan^{-1}\left(\frac{2m_{\rm max}}{h}\right)}_{0}\nonumber\\
&+\left(\frac{k_1e^{-(3k_2)/8}}{4b^2\sqrt{k_2}}\right)\nonumber\\
&\left.\left(\sqrt{\frac{3\pi}{2}}\text{erfi}\left(\frac{\sqrt{k_2}(4\theta^2+3)}{2\sqrt{6}}\right)\right)\right|^{\tan^{-1}\left(\frac{2m_{\rm max}}{h}\right)}_{0}\nonumber\\
&=\zeta_{\rm y},
\end{align}
where $\text{erfi}()$ is the imaginary error function.
For the proofs of \eqref{eq:chnl_Fz2} and \eqref{eq:chnl_Fy2}, see Appendix.

\begin{figure}[!t]
	\centering
	\vspace{-0.0in}
	\includegraphics[width=0.5\textwidth]{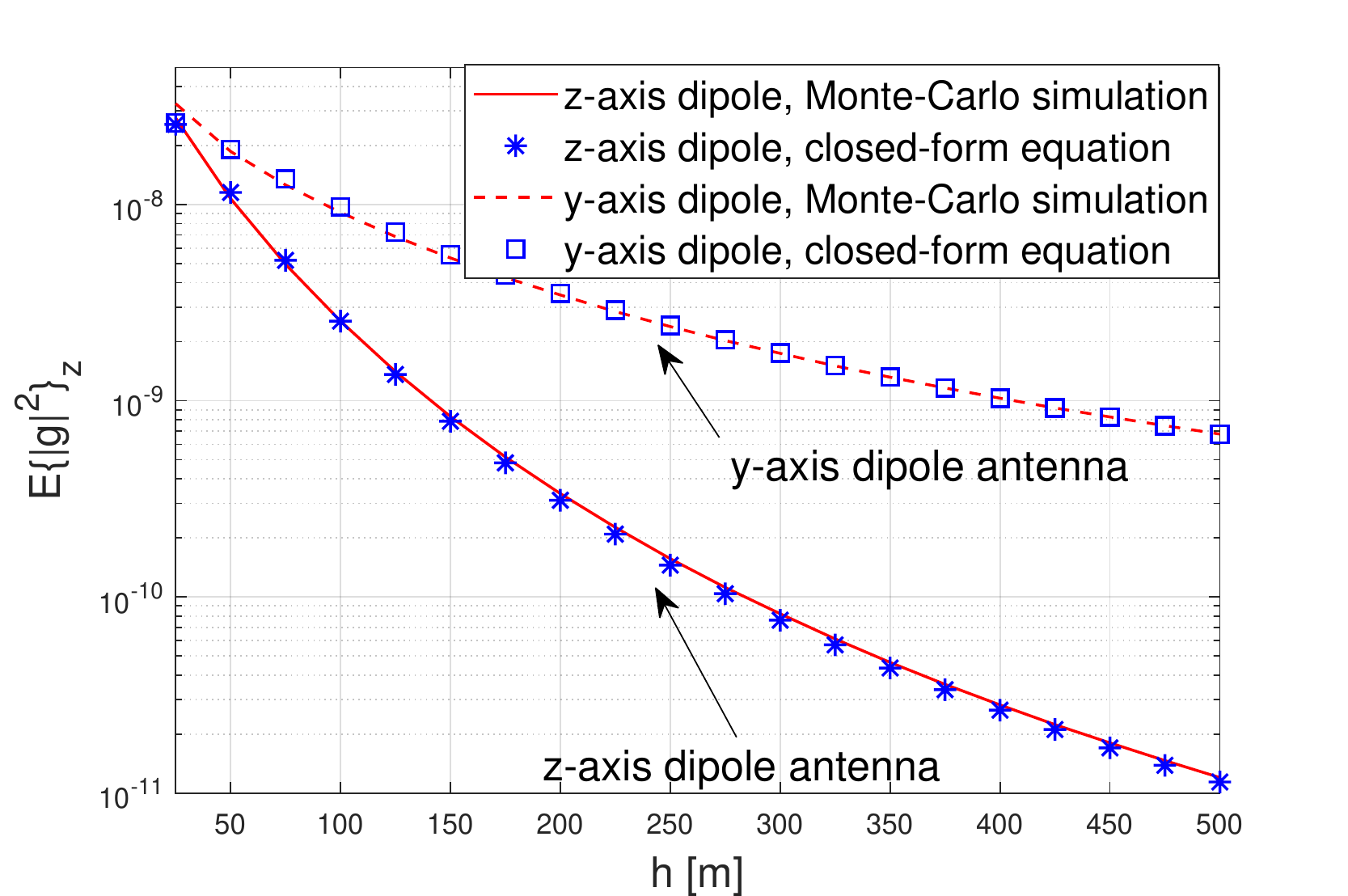}
	\vspace{-0.02in}
	\caption{The expectation of the channel gain verse height ($h$) with different dipole antenna placement on multiple Tx/Rx pairs scenario in \eqref{eq:chnl_Fz2}, \eqref{eq:chnl_Fy2}. $m_0 = 10$, $m_{\rm max} = 100$.}
	\label{fig:chnl_sa2}
	%\vspace{-0.2in}
\end{figure}

The accuracy of the approximations \eqref{eq:chnl_Fz2} and \eqref{eq:chnl_Fy2} are shown in Fig.~\ref{fig:chnl_sa2}. We observe the similar tendency with stand-alone scenario in Fig.~\ref{fig:chnl_sa}. The closed-form equations are close to the exact values and they are closer to the exact values as $h$ grows.

\subsection{The Proposed Scheme}
We propose an antenna selection strategy where the ground transmitters which are connected with the ground receivers utilize $z$-axis dipole antenna, and the ground transmitters which are connected with the aerial receivers utilize $y$-axis dipole antenna. 
In practice, transmitters need to decode messages or receive preambles from receivers in order to know whether the type of receiver is ground or aerial ones. One feasible way to decide the antenna is by measuring received signal power of preambles. Received signals power from the ground receiver will be higher by selecting  $z$-axis dipole antenna, while received signals power from the aerial receiver will be higher by selecting  $y$-axis dipole antenna.
Let $\mathcal{G}=\{F_{\rm z}^2,F_{\rm y}^2\}$ be the candidate antenna selection. Then, cardinality  $|\mathcal{G}|=2$ for cross-dipole setting, either selecting $z$-axis dipole antenna or $y$-axis dipole antenna. The decision made by the highest received preamble signal power:
\begin{align}\label{eq:ant_sel}
    \{G^{\rm{Tx}}_{i}\}^{\boldsymbol{\star}}&=\arg \max_{\forall G^{\rm{Tx}}_{i}\in\mathcal{G}}PG_i^{\rm{Tx}}\beta_{i,i}\alpha^2_{i,i}\sigma^2_{s},
\end{align}
where $\sigma^2_{s}$ is the signal power of the preamble.

The ergodic achievable rate of the aerial receiver ($i_{\rm th} $ receiver) can be expressed from \eqref{eq:rate_apx} as
\begin{align}\label{eq:rate_apx2}
S_i&\approx\log_2\left(1+\frac{\mathbb{E}\{|g_{i,i}|^2\}}{\sum_{j\neq i}\mathbb{E}\{|g_{i,j}|^2\}}\right)\nonumber\\
&=\log_2\left(1+\frac{\zeta_{\rm y}}{(K_{\rm grd})\zeta_{\rm z}+(K_{\rm arl}-1)\zeta_{\rm y}}\right)~,
\end{align}
where $\zeta_{\rm z}$, $\zeta_{\rm y}$ come from \eqref{eq:chnl_Fz2} and \eqref{eq:chnl_Fy2}, and $K_{\rm grd}$, $K_{\rm arl}$ are the number of the grounds transmitters which are connected with the ground receiver and the aerial receiver respectively; $K_{\rm grd}+K_{\rm arl}=K$.

%\vspace{-0.02in}

\section{Numerical Results} \label{sec:results}
In this section, we present simulation results for the performance of the ergodic achievable rate and the sum rate in the proposed schemes. The simulation settings are listed in Table~\ref{table:simulation_settings}. We assume that transmitters perfectly know the type of receivers (ground or aerial devices).
\begin{table}[!h]
\renewcommand{\arraystretch}{1.1}
\caption{Simulation settings}
\label{table:simulation_settings}
\centering
\begin{tabular}{lc}
\hline
Parameter & Value \\
\hline\hline
Minimum radius of circle \\in ground plane ($m_0$)& 10 m \\\hline
Maximum radius of circle \\in ground plane ($m_{\rm max}$)& 100 m \\\hline
The number of Tx/Rx pairs (K)& 5, 10 \\\hline
Transmit power (P) & 23 dBm \\\hline
Carrier frequency ($f_0$) & 800 MHz \\\hline
Bandwidth ($B$)& 200 kHz \\\hline
Additive noise power & $-174 + 10\log_{10}(B)$ dBm\\
\hline
\end{tabular}
\end{table}

\begin{figure}[!t]
	\centering
	\subfloat[Ergodic achievable rate]{
	\vspace{-0.0in}
	\includegraphics[width=0.5\textwidth]{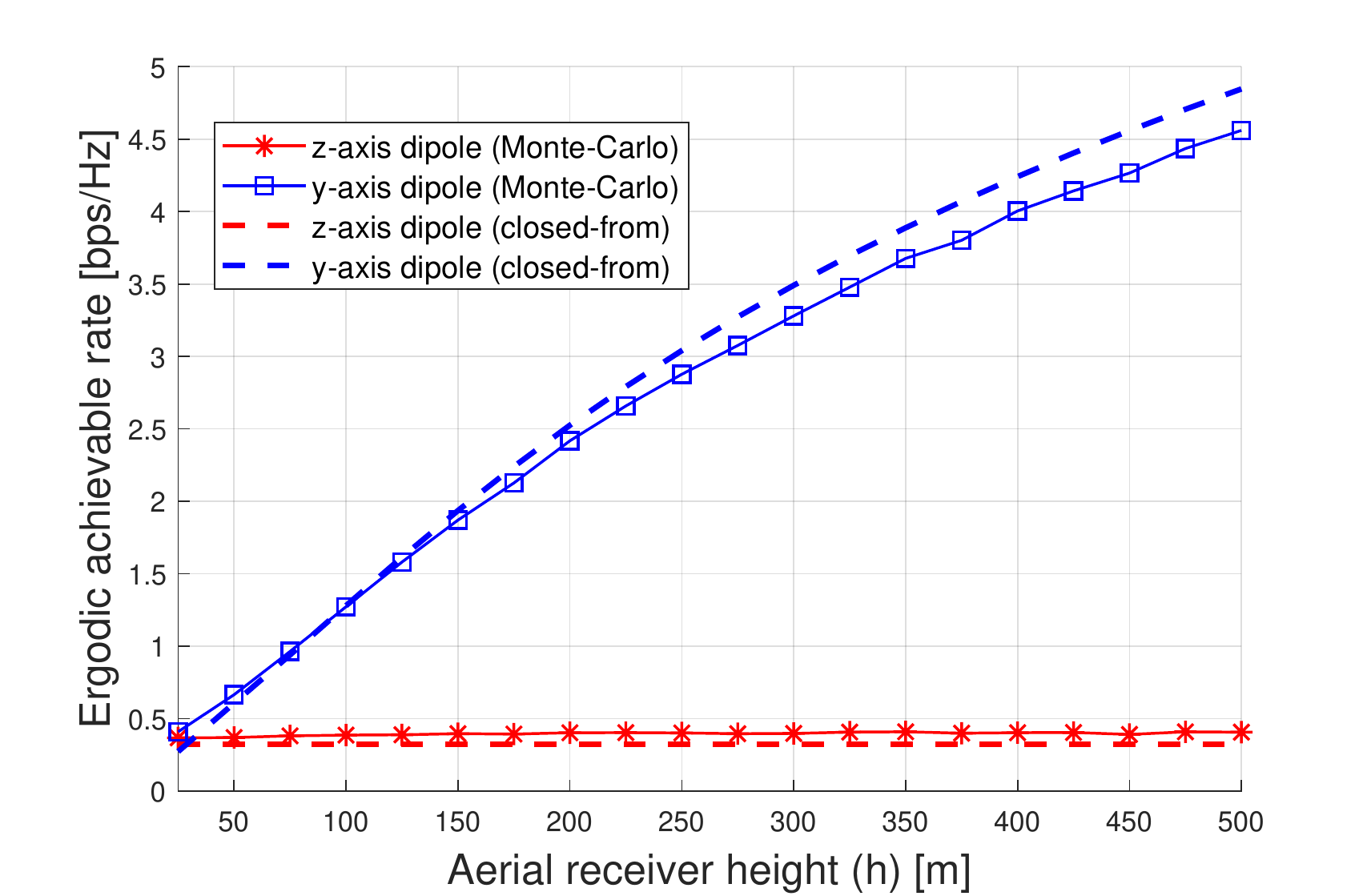}\label{fig:rate_a}}
	\vspace{-0.01in}
	\subfloat[Desired and interference signal power]{
	\includegraphics[width=0.5\textwidth]{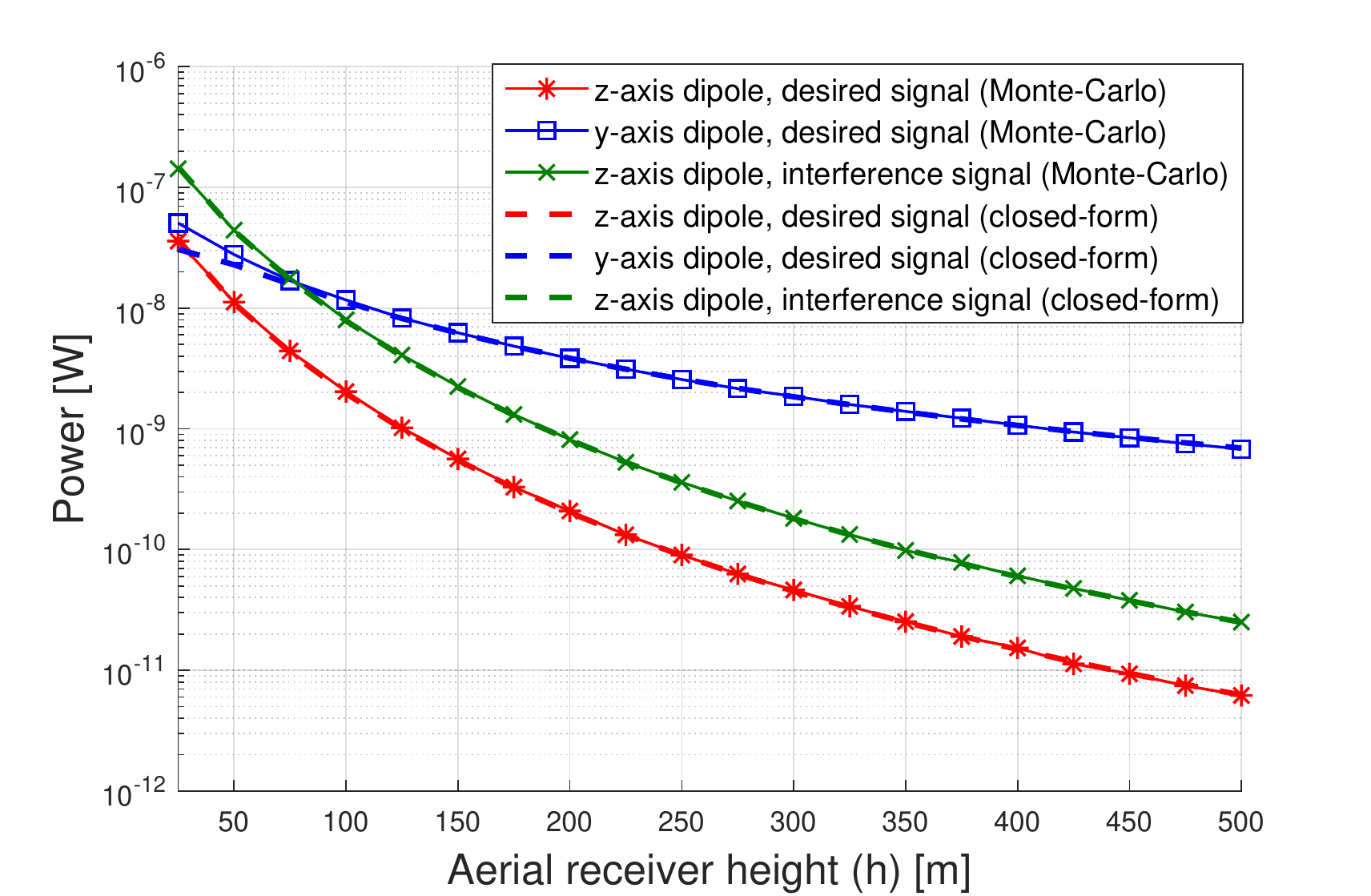}\label{fig:pw_a}}
	\vspace{-0.0in}
	\caption{Ergodic achievable rate and signal power verse the height of the aerial receiver on the scenario in Section \ref{sec:scen1} with $K=5$.} 
	\label{fig:rate_1}
	%\vspace{-0.2in}
\end{figure}

\subsection{Ergodic Achievable Rate of the Aerial Receiver on the Single Rx Scenario}
In Fig.~\ref{fig:rate_1}, we show the ergodic achievable rate, and desired and interference signal power of the aerial receiver on the scenario in Section \ref{sec:scen1}. In the scenario, the single aerial receiver is fixed at $(0,0,h)$ m. In Fig.~\ref{fig:rate_1}\subref{fig:rate_a}, the closed-form equation of $z$-axis dipole antenna (the red dashed line) comes from \eqref{eq:rate_Fz}, and the closed-form equation of $y$-axis dipole antenna (the blue dashed line) comes from \eqref{eq:rate_Fy}. The Monte-Carlo results are directly simulated by the exact ergodic achievable rate in \eqref{eq:rate}. Thus, the gap between closed-form and Monte-Carlo results comes from the approximation procedure in \eqref{eq:rate_apx}. It is observed that the rate gradually increases as the height of the aerial receiver grows in $y$-axis dipole setting from the ground transmitter (blue curve), while the rate is static in $z$-axis dipole setting from the ground transmitter (red curve). The simulation result in Fig.~\ref{fig:rate_1}\subref{fig:pw_a} shows the consistency with Fig.~\ref{fig:rate_1}\subref{fig:rate_a} that the gap between desired signal power of $y$-axis dipole (blue curve) and interference signal power (green curve) keeps increasing, while the gap between desired signal power of $z$-axis dipole (red curve) and interference signal power (green curve) is constant. Note that we include only the result of interference signal power of $z$-axis dipole antenna setting, since both the interference signal power of $z$-axis dipole and $y$-axis dipole is the same. From the simulation result, we conclude that the performance can be improved by the different antenna radiation pattern depending on the type (ground / aerial) of the receiver. In addition, we show that as the height of the aerial receiver increases, the performance becomes better. 

\subsection{Ergodic Achievable Rate of the Aerial Receiver on the Multiple Tx/Rx Pair Scenario}
\begin{figure}[!t]
	\centering
	\vspace{-0.0in}
	\includegraphics[width=0.5\textwidth]{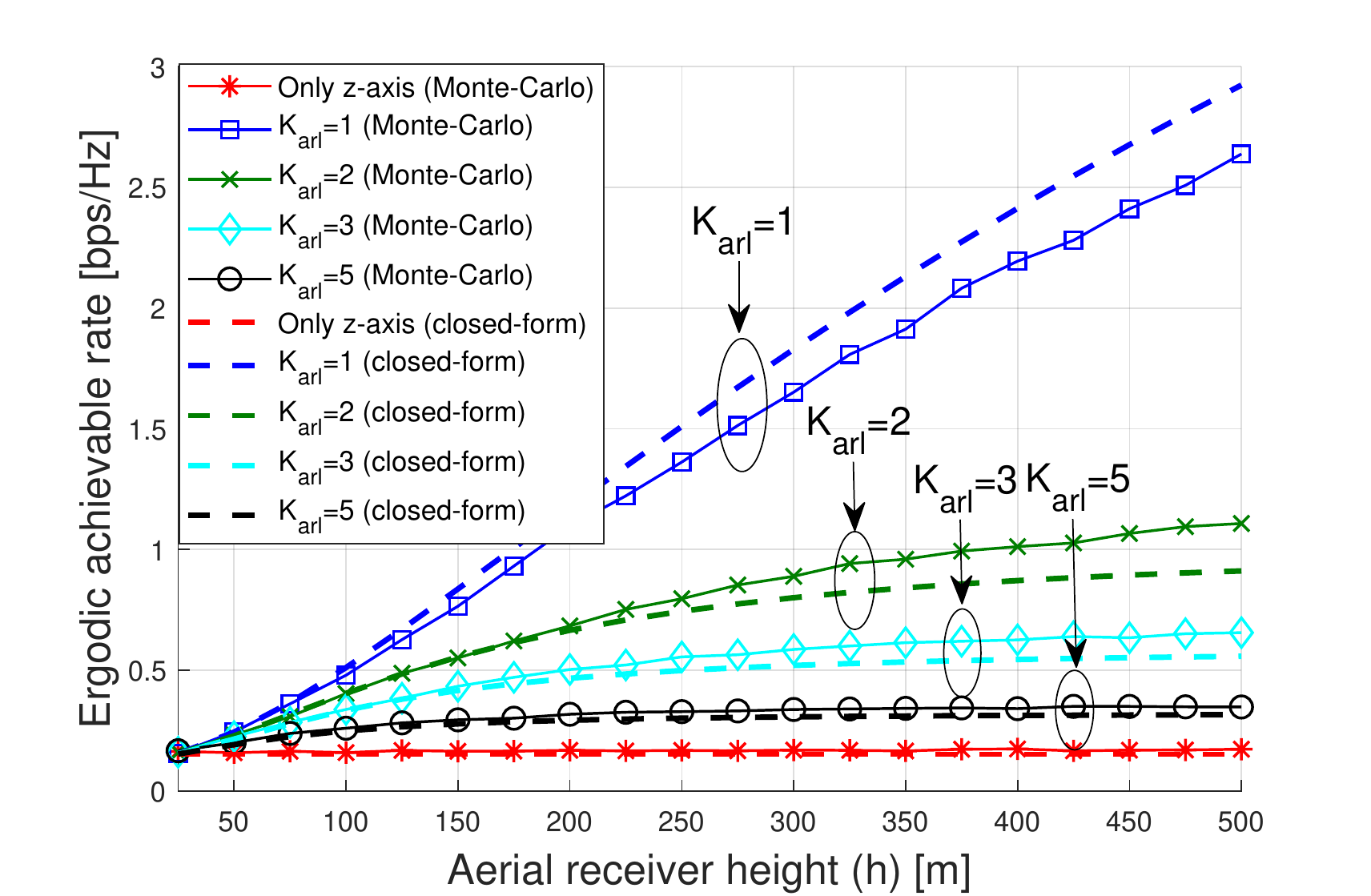}
	\vspace{-0.0in}
	\caption{Ergodic achievable rate verse the height of the aerial receivers on the scenario in Section \ref{sec:scen2} with the different number of aerial receivers ($K_{\rm arl}$), $K=10$.}
	\label{fig:rate_2}
	\vspace{-0.2in}
\end{figure}

The ergodic achievable rate of the  scenario in Section \ref{sec:scen2} is shown in Fig.~\ref{fig:rate_2}. In the scenario, all the ground / aerial devices are randomly located at the given space. The transmitters which are connected with the ground receiver utilize $z$-axis dipole antenna, and the transmitters which are connected with the aerial receiver utilize $y$-axis dipole antenna. The number of Tx/Rx pairs ($K$) is $10$. The simulation results of closed-form equation come from \eqref{eq:rate_apx2} with the different number of aerial receivers ($K_{\rm arl}$). We observe that the performance of the achievable rate increases as the height of the aerial receiver increases, while the performance decreases as the number of the aerial receivers increases. It means that as the number of aerial receivers increases, the number of transmitters which utilize $y$-axis dipole antenna proportionally increases. Then, the power of interference signal toward individual aerial receivers also increases, which degrades the performance. As a reference, we include the simulation results that all transmitters utilize $z$-axis dipole antenna (red curve), which achieves worse performance with the flat curve.

\begin{figure}[!t]
	\centering
	\vspace{-0.0in}
	\includegraphics[width=0.5\textwidth]{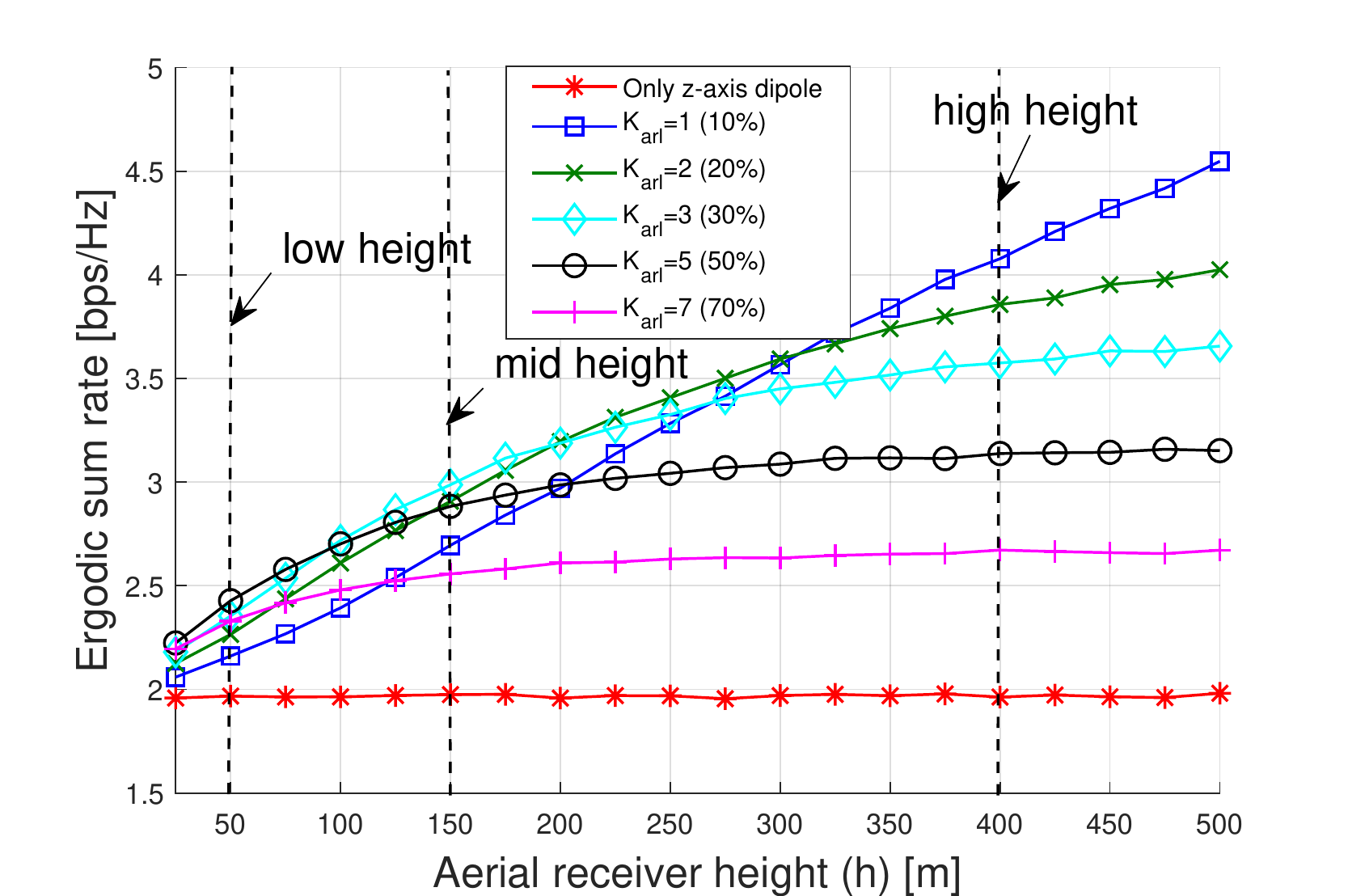}
	\vspace{-0.0in}
	\caption{Ergodic sum rate verse the height of the aerial receivers on the scenario in Section \ref{sec:scen2} with the different number of aerial receivers ($K_{\rm arl}$), $K=10$.}
	\label{fig:sumrate}
	\vspace{-0.2in}
\end{figure}

\begin{figure}[!t]
	\centering
	\vspace{-0.0in}
	\includegraphics[width=0.5\textwidth]{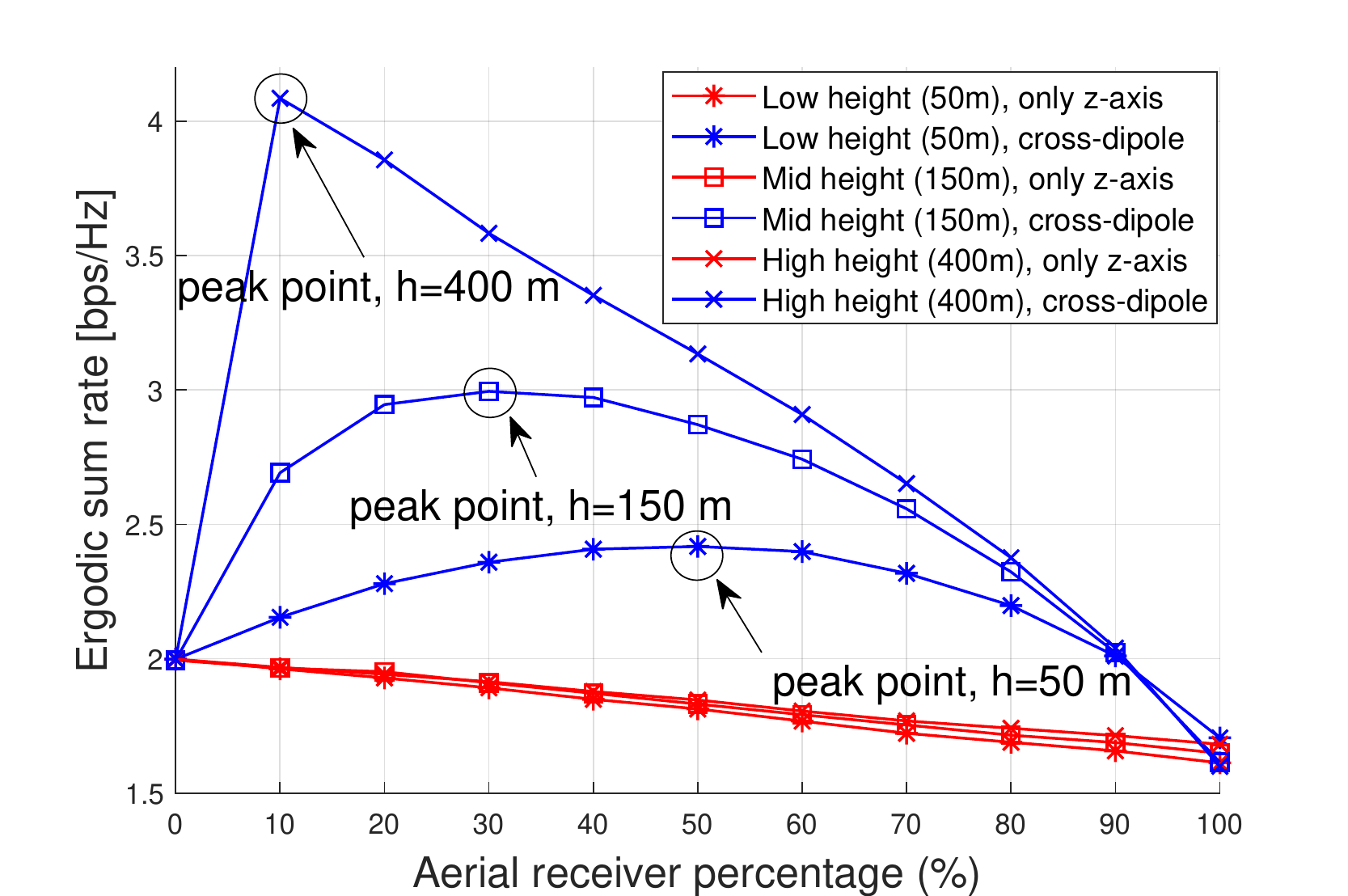}
	\vspace{-0.1in}
	\caption{Ergodic sum rate verse the percentage of the aerial receivers in the network with the different height of the aerial receivers, $K=10$.}
	\label{fig:sumrate_h}
	%\vspace{-0.2in}
\end{figure}

\begin{figure}[!t]
	\centering
	\vspace{-0.0in}
	\includegraphics[width=0.5\textwidth]{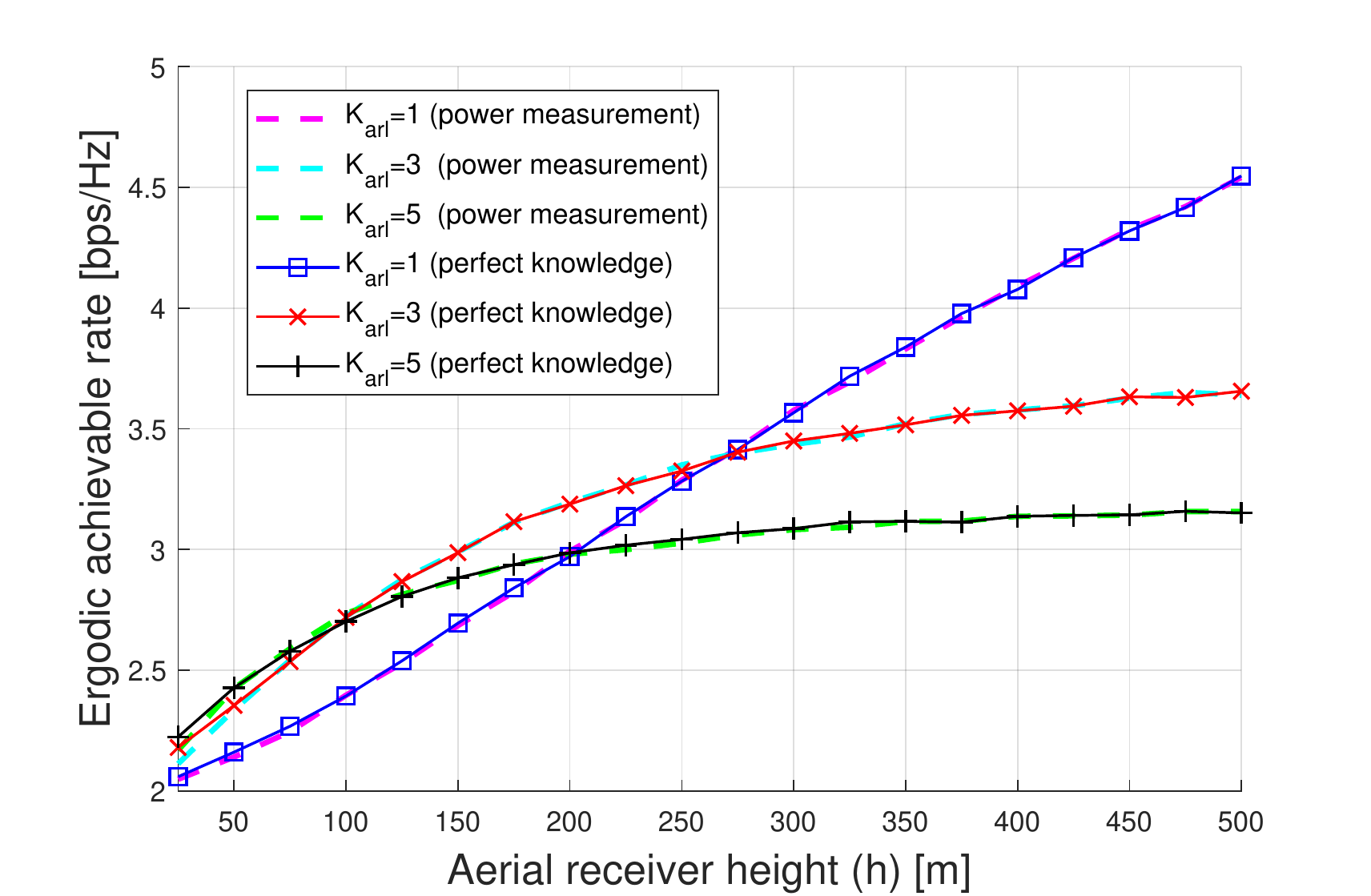}
	\vspace{-0.1in}
	\caption{Ergodic sum rate verse the height of aerial receivers with antenna selections based on perfect knowledge and power measurement, $K=10$.}
	\label{fig:ant_sel}
	%\vspace{-0.2in}
\end{figure}

\begin{figure}[!t]
	\centering
	\vspace{-0.0in}
	\includegraphics[width=0.5\textwidth]{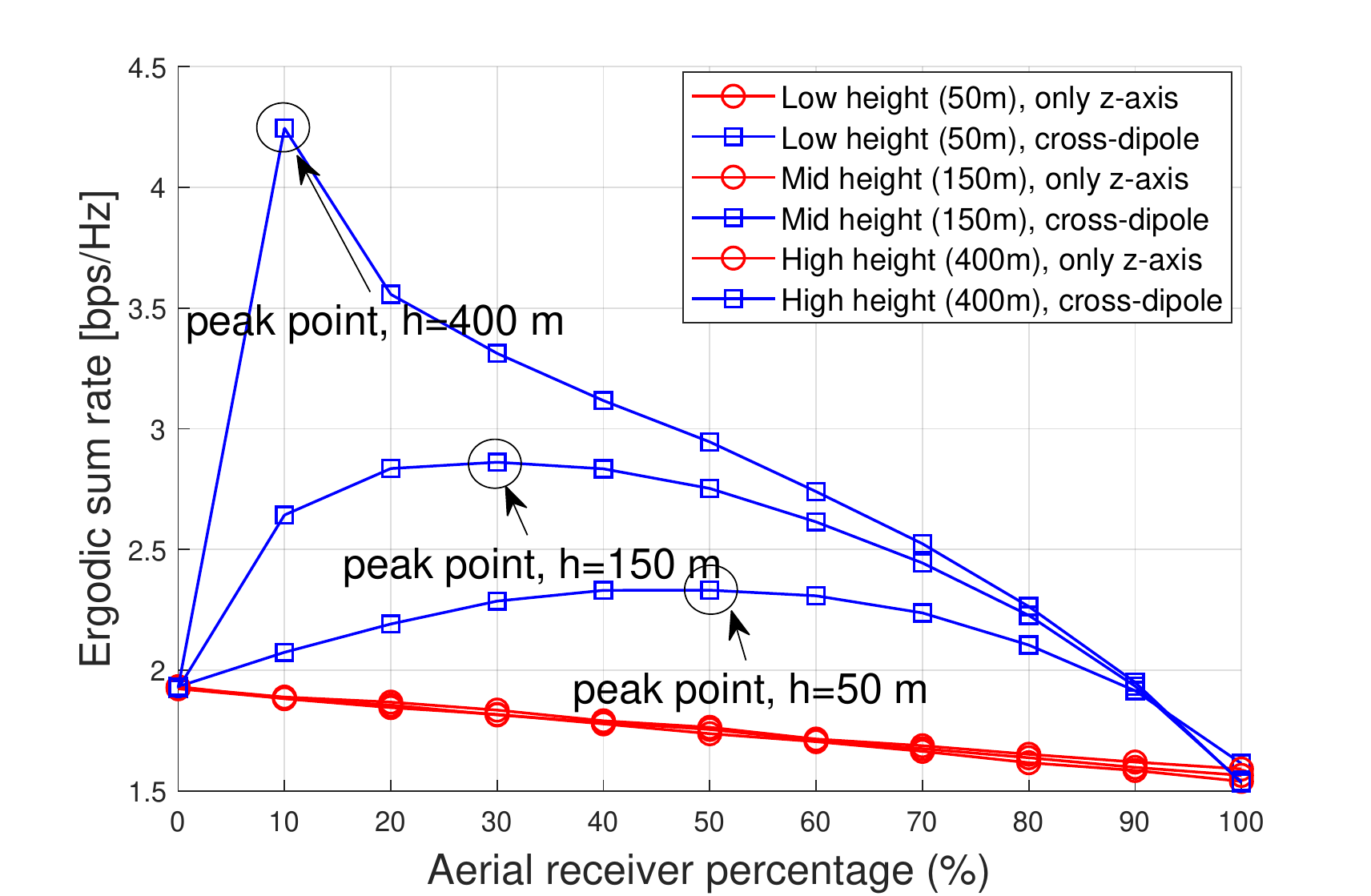}
	\vspace{-0.1in}
	\caption{Ergodic sum rate verse the percentage of the aerial receivers in the network by Rician fading channel in \eqref{eq:chnl_ri}, where K-factor $=10$ dB.}
	\label{fig:sum_rate_ri}
	%\vspace{-0.2in}
\end{figure}

\subsection{Ergodic Sum Rate of the Aerial Receiver on the Multiple Tx/Rx Pair Scenario}
In Fig.~\ref{fig:sumrate}, we depict the ergodic sum rate on scenario in Section \ref{sec:scen2} by Monte-Carlo simulations. The number of Tx/Rx pairs is 10 ($K=10$). the ergodic sum rate is given by $S_{\rm sum}=\sum_{i=1}^K S_i$. If we increase the number of the aerial receivers ($K_{\rm arl}$), the percentage of the aerial receivers in the network proportionally increases. For example, $K_{\rm arl}=1$ means that 10\% of the overall network are the aerial receivers, while $K_{\rm arl}=5$ means that 50\% of the network are the aerial receivers. It is observed that depending on the height of aerial receivers, the best ratio between the number of ground and aerial receivers varies. In the case of the low height ($h=50$ m), 50\% of the aerial receivers ($K_{\rm arl}=5$) in the network achieves the best rate (black circle line) and the 10\% of the aerial receivers ($K_{\rm arl}=1$) in the network achieves the worst rate (blue square line). On the other hand, in case of the medium height ($h=150$ m), 30\% of the aerial receivers in the network achieves the best rate (cyan diamond line) and 70\% of the aerial receivers in the network achieves the worst rate (purple plus line). Furthermore, in the case of the high height ($h=400$ m), 10\% of the aerial receivers in the network achieves the best rate (blue square line) and 70\% of the aerial receivers in the network achieves the worst rate (purple plus line). Note that the $y$-axis dipole antenna to the aerial receiver improves the achievable rate of not only the aerial receivers but also the ground receivers, which reduces the interference in both case.

The ergodic sum rate verse the percentage of the aerial receivers in the network with different height is shown in Fig.~\ref{fig:sumrate_h}. We can observe that the point of the peak rate on the percentage of the aerial receivers is different depending on the height. The 50 \% / 30\% / 10\% of the aerial receivers is the peak in low / medium / high height of the aerial receivers. Note that this observation coincides with the result discussed in Fig.~\ref{fig:sumrate}. We conclude that the best ratio of the ground / aerial receivers in the network is changed by the height of the aerial receivers. In low height, 50\% aerial nodes can be the best, but the single aerial receivers can be the best in the high height of the aerial receivers.
Also, we compare the proposed cross-dipole antenna scheme (blue curves) with the scheme that transmitters use only $z$-axis dipole antenna (red curves). We observe that our proposed scheme achieves better performance in all heights.

\subsection{The Effect of Antenna Selection based on Received Signal Power Measurement}
In previous simulation results, we assume that transmitters know information of the type of receivers, and decide $z$-axis or $y$-axis dipole antenna without estimation. In this part, we apply the strategy in \eqref{eq:ant_sel} that selects antenna by measuring signal power. In Fig.~\ref{fig:ant_sel}, we compare the ergodic sum rate of antenna selection strategies based on perfect knowledge and measurement. It is observed that the performance between two strategies is close to each other, which means that the loss of performance is trivial by the antenna selection based on signal power. 

\subsection{The Performance on Rician fading Channel Model}\label{sec:rician}
In Fig.~\ref{fig:sum_rate_ri}, we show the performance in Rician fading channel model in \eqref{eq:chnl_ri} by the ergodic sum rate. We observe that a tendency for the ergodic sum rate is similar to Rayleigh fading channel model in \eqref{eq:chnl}. Note that since the expectation value of the small scale fading coefficient is the same for both Rayleigh and Rician fading model, the analytical approximation result of ergodic achievable rate is the same; $\mathbb{E}\left\{\alpha^2\right\}=\mathbb{E}\left\{\frac{\kappa}{\kappa+1}+\frac{1}{\kappa+1}\alpha^2\right\}=1$.

\section{Conclusion}
\label{sec:conclusion}
In this paper, we propose and study an interference mitigation scheme that utilizes the diversity of the radiation pattern in 3D topology IoT uncoordinated network. We propose a cross-dipole antenna setting at the transmitter, which utilizes either $z$-axis dipole antenna or $y$-axis dipole antenna depending on the type of the receivers. We design the 3D topology channel model based on the location of the devices, and we analytically show that $y$-axis dipole antenna achieves a better performance for aerial receivers than $z$-axis dipole antenna. By simulation results, we observe that cross-dipole antenna scheme outperforms the scheme that only uses the $z$-dipole antenna, and we can improve the performance by increasing the height of the aerial receivers. In addition, depending on the height of the aerial receivers (low / mid / high), the best ratio of the ground receiver to aerial receiver varies.
\vspace{-0.1in}

\appendix[Proof of Equations \eqref{eq:chnl_Fz2}, \eqref{eq:chnl_Fy2}]
We may rewrite \eqref{eq:exp_Fz} as:
\begin{align}\label{}
&\mathbb{E}\{|g_{i,i}|^2\}_{\rm z}\nonumber\\
&=k_1\mathbb{E}\left\{\frac{(F_{\rm{z}})^2}{R^2}\right\}\nonumber\\
&=k_1\mathbb{E}\left\{\left(\frac{\cos\left(\frac{\pi}{2}\cos\theta\right)}{\sin\theta}\right)^2\frac{\cos^2\theta}{h^2}\right\}\nonumber\\
&=k_1\int\left(\frac{\cos\left(\frac{\pi}{2}\cos\theta\right)}{h\sin\theta}\right)^2\cos^2\theta\ f_\theta(\theta){\rm d}\theta\nonumber\\
&=\frac{k_1}{b^2}\int^{\tan^{-1}\left(\frac{2m_{\rm max}}{h}\right)}_{0}\left(\frac{\cos\left(\frac{\pi}{2}\cos\theta\right)}{\sin\theta}\right)^2\left(\tan\theta e^{k_2\tan^2\theta}\right){\rm d}\theta~,
\end{align}
where $F_{\rm{z}}$, $f_\theta(\theta)$ come from \eqref{eq:F_z re}, \eqref{eq:pdf_th2}, and $k_2=\frac{-h^2}{2b^2}$. If the height ($h$) goes to infinity, the interval of  the integral with respect to $\theta$ goes to $0$ ($\tan^{-1}(\frac{2m_{\rm max}}{h})\to0, \text{ as } h\to\infty$). Then, we can apply Taylor series approximation at $\theta=0$ as follows:
\begin{align}\label{eq:exp_Fz2}
&\mathbb{E}\{|g_{i,i}|^2\}_{\rm z}\nonumber\\
&\approx\frac{k_1}{b^2}\int^{\tan^{-1}\left(\frac{2m_{\rm max}}{h}\right)}_{0}\left(\frac{\pi^2\theta^3}{16}\right)\left(e^{k_2(\theta^2+\frac{2}{3}\theta^4)}\right){\rm d}\theta,
\end{align}
where $\frac{\pi^2\theta^3}{16}$ is the first term of Taylor series of $(\cos (\frac{\pi}{2}\cos\theta)$ $/\sin\theta)^2$$(\tan\theta)$, and $\theta^2+\frac{2}{3}\theta^4$ includes the first and the second terms of the Taylor series of $\tan^2\theta$. By solving \eqref{eq:exp_Fz2}, we can obtain \eqref{eq:chnl_Fz2}.

Similarly, the proof of equation \eqref{eq:chnl_Fy2} is obtained by starting from \eqref{eq:EgFy} as follows:
\begin{align}\label{eq:exp_Fy2}
&\mathbb{E}\{|g_{i,i}|^2\}_{\rm y}\nonumber\\
&=k_1\mathbb{E}\left\{\frac{(F_{\rm{y}})^2}{R^2}\right\}\nonumber\\
&=k_1\mathbb{E}\left\{\left(\frac{\cos\left(\frac{\pi}{2}\cos(\cos^{-1}(\sin(\theta)\sin(\hat{\phi})))\right)}{h\sin\left(\cos^{-1}(\sin(\theta)\sin(\hat{\phi}))\right)}\right)^2\frac{\cos^2\theta}{h^2}\right\}\nonumber\\
&=k_1\int\int\left(\frac{\cos\left(\frac{\pi}{2}\cos(\cos^{-1}(\sin(\theta)\sin(\hat{\phi})))\right)}{h\sin\left(\cos^{-1}(\sin(\theta)\sin(\hat{\phi}))\right)}\right)^2\nonumber\nonumber\\
&\qquad\times\cos^2\theta\ f_\theta(\theta)f_{\hat{\phi}}(\hat{\phi}){\rm d}\theta {\rm d}\hat{\phi}\nonumber\\
&=\frac{k_1}{2\pi^2b^2 }\int^{2\pi}_0\int^{\tan^{-1}\left(\frac{2m_{\rm max}}{h}\right)}_{0}\nonumber\\
&\qquad\times\left(\frac{\cos\left(\frac{\pi}{2}\cos(\cos^{-1}(\sin(\theta)\sin(\hat{\phi})))\right)}{\sin\left(\cos^{-1}(\sin(\theta)\sin(\hat{\phi}))\right)}\right)^2\nonumber\\
&\qquad\qquad\times\left(\tan\theta e^{k_2\tan^2\theta}\right)\left(2\pi-\hat{\phi}\right){\rm d}\theta {\rm d}\hat{\phi}~,
\end{align}
where $F_{\rm{y}}$, $f_{\hat{\phi}}(\hat{\phi})$, and $f_\theta(\theta)$ come from \eqref{eq:F_y}, \eqref{eq:pdf_phi2}, \eqref{eq:pdf_th2}, respectively. After  applying Taylor series approximation, \eqref{eq:exp_Fy2} can be rewritten  as
\begin{align}\label{eq:exp_Fy_apx2}
&\mathbb{E}\{|g_{i,i}|^2\}_{\rm y}\nonumber\\
&\approx\frac{k_1}{2\pi^2b^2}\int^{2\pi}_0\int^{\tan^{-1}\left(\frac{2m_{\rm max}}{h}\right)}_{0}\nonumber\\
&\qquad\times\left(\theta+\left(\sin^2(\hat{\phi})-\frac{\pi^2\sin^2(\hat{\phi})}{4}+\frac{1}{3}\right)\theta^3\right)\nonumber\\
&\qquad\qquad\times\left(e^{k_2(\theta^2+\frac{2}{3}\theta^4)}\right)\left(2\pi-\hat{\phi}\right){\rm d}\theta {\rm d}\hat{\phi}\nonumber\\
&=\frac{k_1}{2\pi^2b^2 }\int^{\tan^{-1}\left(\frac{2m_{\rm max}}{h}\right)}_{0}\left(2\pi^2\theta+\left(\frac{5\pi^2}{3}-\frac{\pi^4}{4}\right)\theta^3\right)\nonumber\\
&\qquad\times\left(e^{k_2(\theta^2+\frac{2}{3}\theta^4)}\right){\rm d}\theta,
\end{align}
where $\left(\theta+\left(\sin^2(\hat{\phi})-\frac{\pi^2\sin^2(\hat{\phi})}{4}+\frac{1}{3}\right)\theta^3\right)$ includes the first and the second terms of the Taylor series at $\theta=0$ with fixed $\hat{\phi}$. By solving \eqref{eq:exp_Fy_apx2}, we can obtain \eqref{eq:chnl_Fy2}.

\bibliographystyle{IEEEtran} 
\bibliography{IEEEabrv,bibfile}

\begin{thebibliography}{10}
\providecommand{\url}[1]{#1}
\csname url@samestyle\endcsname
\providecommand{\newblock}{\relax}
\providecommand{\bibinfo}[2]{#2}
\providecommand{\BIBentrySTDinterwordspacing}{\spaceskip=0pt\relax}
\providecommand{\BIBentryALTinterwordstretchfactor}{4}
\providecommand{\BIBentryALTinterwordspacing}{\spaceskip=\fontdimen2\font plus
\BIBentryALTinterwordstretchfactor\fontdimen3\font minus
  \fontdimen4\font\relax}
\providecommand{\BIBforeignlanguage}[2]{{%
\expandafter\ifx\csname l@#1\endcsname\relax
\typeout{** WARNING: IEEEtran.bst: No hyphenation pattern has been}%
\typeout{** loaded for the language `#1'. Using the pattern for}%
\typeout{** the default language instead.}%
\else
\language=\csname l@#1\endcsname
\fi
#2}}
\providecommand{\BIBdecl}{\relax}
\BIBdecl

\bibitem{maeng2019interference}
S.~J. Maeng, M.~A. Deshmukh, I.~Guvenc, and A.~Bhuyan, ``Interference
  mitigation scheme in {3D} topology {IoT} network with antenna radiation
  pattern,'' in \emph{Proc. IEEE Veh. Technol. Conf. (VTC)}, Honolulu, HI, Sep.
  2019, pp. 1--6.

\bibitem{gupta2015survey}
A.~Gupta and R.~K. Jha, ``A survey of {5G} network: Architecture and emerging
  technologies,'' \emph{IEEE Access}, vol.~3, pp. 1206--1232, Jul. 2015.

\bibitem{whitmore2015internet}
A.~Whitmore, A.~Agarwal, and L.~Da~Xu, ``The {Internet} of things—a survey of
  topics and trends,'' \emph{Information Systems Frontiers}, vol.~17, no.~2,
  pp. 261--274, Apr. 2015.

\bibitem{ijaz2016enabling}
A.~Ijaz, L.~Zhang, M.~Grau, A.~Mohamed, S.~Vural, A.~U. Quddus, M.~A. Imran,
  C.~H. Foh, and R.~Tafazolli, ``Enabling massive {IoT} in {5G} and beyond
  systems: Phy radio frame design considerations,'' \emph{IEEE Access}, vol.~4,
  pp. 3322--3339, June 2016.

\bibitem{al2015internet}
A.~Al-Fuqaha, M.~Guizani, M.~Mohammadi, M.~Aledhari, and M.~Ayyash, ``Internet
  of things: A survey on enabling technologies, protocols, and applications,''
  \emph{IEEE Commun. Surveys Tuts.}, vol.~17, no.~4, pp. 2347--2376, June 2015.

\bibitem{popli2018survey}
S.~Popli, R.~K. Jha, and S.~Jain, ``A survey on energy efficient narrowband
  internet of things ({NBIoT}): architecture, application and challenges,''
  \emph{IEEE Access}, vol.~7, pp. 16\,739--16\,776, Nov. 2018.

\bibitem{rico2016overview}
A.~Rico-Alvarino, M.~Vajapeyam, H.~Xu, X.~Wang, Y.~Blankenship, J.~Bergman,
  T.~Tirronen, and E.~Yavuz, ``An overview of {3GPP} enhancements on machine to
  machine communications,'' \emph{IEEE Commun. Mag.}, vol.~54, no.~6, pp.
  14--21, June 2016.

\bibitem{sinha2017survey}
R.~S. Sinha, Y.~Wei, and S.-H. Hwang, ``A survey on {LPWA} technology: {LoRa}
  and {NB-IoT},'' \emph{ICT Express}, vol.~3, no.~1, pp. 14--21, Mar. 2017.

\bibitem{zeng2018cellular}
Y.~Zeng, J.~Lyu, and R.~Zhang, ``Cellular-connected {UAV}: Potential,
  challenges, and promising technologies,'' \emph{IEEE Wireless Commun.},
  vol.~26, no.~1, pp. 120--127, Sep. 2018.

\bibitem{xiao2016enabling}
Z.~Xiao, P.~Xia, and X.-G. Xia, ``Enabling {UAV} cellular with millimeter-wave
  communication: Potentials and approaches,'' \emph{IEEE Commun. Mag.},
  vol.~54, no.~5, pp. 66--73, May 2016.

\bibitem{motlagh2017uav}
N.~H. Motlagh, M.~Bagaa, and T.~Taleb, ``{UAV}-based {IoT} platform: A crowd
  surveillance use case,'' \emph{IEEE Commun. Mag.}, vol.~55, no.~2, pp.
  128--134, Feb. 2017.

\bibitem{wei2016gyro}
T.~Wei and X.~Zhang, ``Gyro in the air: tracking {3D} orientation of
  batteryless {I}nternet-of-things,'' in \emph{Proc. Int. Conf. Mobile
  Computing and Networking}, 2016, pp. 55--68.

\bibitem{ciftler2017iot}
B.~S. Ciftler, A.~Kadri, and I.~G{\"u}ven{\c{c}}, ``{IoT} localization for
  bistatic passive {UHF RFID} systems with {3-D} radiation pattern,''
  \emph{IEEE Internet of Things J.}, vol.~4, no.~4, pp. 905--916, Apr. 2017.

\bibitem{bujari2018standards}
A.~Bujari, M.~Furini, F.~Mandreoli, R.~Martoglia, M.~Montangero, and
  D.~Ronzani, ``Standards, security and business models: key challenges for the
  {IoT} scenario,'' \emph{Mobile Networks and Applications}, vol.~23, no.~1,
  pp. 147--154, Feb. 2018.

\bibitem{teng2018resource}
Y.~Teng, M.~Liu, F.~R. Yu, V.~C. Leung, M.~Song, and Y.~Zhang, ``Resource
  allocation for ultra-dense networks: A survey, some research issues and
  challenges,'' \emph{IEEE Commun. Surveys Tuts.}, Aug. 2018.

\bibitem{zucchetto2017uncoordinated}
D.~Zucchetto and A.~Zanella, ``Uncoordinated access schemes for the {IoT}:
  approaches, regulations, and performance,'' \emph{IEEE Commun. Mag.},
  vol.~55, no.~9, pp. 48--54, Sep. 2017.

\bibitem{kammoun20153d}
A.~Kammoun, M.~Debbah, M.-S. Alouini \emph{et~al.}, ``{3D massive MIMO}
  systems: Modeling and performance analysis,'' \emph{IEEE Transactions on
  wireless communications}, vol.~14, no.~12, pp. 6926--6939, 2015.

\bibitem{baianifar2017impact}
M.~Baianifar, S.~Khavari, S.~M. Razavizadeh, and T.~Svensson, ``Impact of user
  height on the coverage of {3D} beamforming-enabled massive {MIMO} systems,''
  in \emph{Proc. Int. Symp. Personal, Indoor, and Mobile Radio Commun.
  (PIMRC)}, 2017, pp. 1--5.

\bibitem{li2013dynamic}
Y.~Li, X.~Ji, D.~Liang, and Y.~Li, ``Dynamic beamforming for three-dimensional
  mimo technique in lte-advanced networks,'' \emph{International journal of
  antennas and propagation}, vol. 2013, 2013.

\bibitem{rebato2019stochastic}
M.~Rebato, J.~Park, P.~Popovski, E.~De~Carvalho, and M.~Zorzi, ``Stochastic
  geometric coverage analysis in {mmWave} cellular networks with realistic
  channel and antenna radiation models,'' \emph{IEEE Trans. Commun.}, vol.~67,
  no.~5, pp. 3736--3752, Jan. 2019.

\bibitem{3gpp}
3GPP, ``Study on channel model for frequencies from 0.5 to 100 {GHz (Release
  14)},'' 3GPP, Technical Report 38.901, 2017.

\bibitem{geraci2018understanding}
G.~Geraci, A.~Garcia-Rodriguez, L.~G. Giordano, D.~L{\'o}pez-P{\'e}rez, and
  E.~Bj{\"o}rnson, ``Understanding {UAV} cellular communications: from existing
  networks to massive {MIMO},'' \emph{IEEE Access}, vol.~6, pp.
  67\,853--67\,865, Nov. 2018.

\bibitem{zhu20193}
L.~Zhu, J.~Zhang, Z.~Xiao, X.~Cao, D.~O. Wu, and X.-G. Xia, ``{3-D} beamforming
  for flexible coverage in millimeter-wave {UAV} communications,'' \emph{IEEE
  Wireless Commun. Lett.}, vol.~8, no.~3, pp. 837--840, 2019.

\bibitem{chen2018impact}
J.~Chen, D.~Raye, W.~Khawaja, P.~Sinha, and I.~Guvenc, ``Impact of {3D UWB}
  antenna radiation pattern on {Air-to-Ground} drone connectivity,'' in
  \emph{Proc. IEEE Vehic. Technol. Conf. (VTC)}, Sep. 2018, pp. 1--5.

\bibitem{shafi2006polarized}
M.~Shafi, M.~Zhang, A.~L. Moustakas, P.~J. Smith, A.~F. Molisch, F.~Tufvesson,
  and S.~H. Simon, ``Polarized {MIMO} channels in {3-D}: models, measurements
  and mutual information,'' \emph{IEEE J. Sel. Areas Commun.}, vol.~24, no.~3,
  pp. 514--527, Mar. 2006.

\bibitem{dao20113d}
M.-T. Dao, V.-A. Nguyen, Y.-T. Im, S.-O. Park, and G.~Yoon, ``{3D} polarized
  channel modeling and performance comparison of {MIMO} antenna configurations
  with different polarizations,'' \emph{IEEE Trans. Antennas Propag.}, vol.~59,
  no.~7, pp. 2672--2682, May 2011.

\bibitem{zhang2014power}
Q.~Zhang, S.~Jin, K.-K. Wong, H.~Zhu, and M.~Matthaiou, ``Power scaling of
  uplink massive {MIMO} systems with arbitrary-rank channel means,'' \emph{IEEE
  J. Sel. Topics Signal Process}, vol.~8, no.~5, pp. 966--981, May 2014.

\bibitem{fan2015uplink}
L.~Fan, S.~Jin, C.-K. Wen, and H.~Zhang, ``Uplink achievable rate for massive
  {MIMO} systems with low-resolution {ADC},'' \emph{IEEE Commun. Lett.},
  vol.~19, no.~12, pp. 2186--2189, Oct. 2015.

\bibitem{balanis2016antenna}
C.~A. Balanis, \emph{Antenna theory: analysis and design}.\hskip 1em plus 0.5em
  minus 0.4em\relax John wiley \& sons, 2016.

\bibitem{chandhar2017massive}
P.~Chandhar, D.~Danev, and E.~G. Larsson, ``Massive {MIMO} for communications
  with drone swarms,'' \emph{IEEE Trans. Wireless Commun.}, vol.~17, no.~3, pp.
  1604--1629, Dec. 2017.

\end{thebibliography}

\end{document}